\documentclass[useAMS,usenatbib,usegraphicx]{mn2e}
\usepackage{color}
\title[Solar-like oscillations from the depths of the red-giant star KIC\,4351319]
{Solar-like oscillations from the depths of the red-giant star KIC\,4351319 observed with {\it Kepler}}
\author[M.P. Di Mauro et al.]
{M. P. Di Mauro$^1$, D. Cardini$^{1}$,  G.~Catanzaro$^2$, R. Ventura$^2$, C. Barban$^3$,
\newauthor T. R.~Bedding$^4$, 
J. Christensen-Dalsgaard$^5$, J. De Ridder$^6$, S. Hekker$^{7, 8}$, 
\newauthor D. Huber$^4$, T. Kallinger$^{9, 10}$,
A. Miglio$^{11}$, J. Montalban$^{11}$, B. Mosser$^3$,
 D. Stello$^4$, \newauthor K. Uytterhoeven$^{12, 13}$, K. Kinemuchi$^{14}$, H. Kjeldsen$^{5}$,
F. Mullally$^{15}$, M. Still$^{14}$\\
$^1$INAF - IASF, Istituto di Astrofisica Spaziale e Fisica Cosmica, Via del Fosso del Cavaliere 100, 00133 Roma,
Italy\\
$^2$INAF - Osservatorio Astrofisico di Catania, Via S. Sofia 78, 95123 Catania, Italy\\
$^3$LESIA, CNRS, Universit\'e Pierre et Marie Curie, Universit\'e Denis, Diderot, Observatoire de Paris, 92195 Meudon Cedex, France\\
$^4$Sydney Institute for Astronomy (SIfA), School of Physics, University of Sydney , NSW 2006, Australia\\
$^5$Institut for Fysik og Astronomi, Bygn. 1520, Aarhus Universitet, Ny Munkegade, DK-8000 Aarhus C, Denmark\\
$^6$Institut voor Sterrenkunde, K. U. Leuven, Celestijnenlaan 200D, 3001 Leuven, Belgium\\
$^7$School of Physics and Astronomy, University of Birmingham, Birmingham B 15 2TT, UK\\
$^8$Astronomical Institute 'Anton Pannekoek', University of Amsterdam,
Science Park 904,
1098 XH Amsterdam, The Netherlands\\
$^9$Department of Physics and Astronomy, University of British Columbia, Vancouver, Canada\\
$^{10}$Institute of Astronomy, University of Vienna, 1180 Vienna, Austria\\
$^{11}$Institut d'Astrophysique et de G\'eofisique de l'Universit\`e de Li\`ege, All\`e du 6 Aout 17-B 4000 Li\`ege, Belgium\\
$^{12}$Laboratoire AIM, CEA/DSM-CNRS-Universit\'e Paris Diderot; CEA, IRFU, SAp, Centre de Saclay, 91191, Gif-sur-Yvette, France\\
$^{13}$Kiepenheuer-Institut f\"ur Sonnenphysik, Sch\"oneckstra\ss{}e 6, 79104 Freiburg im
Breisgau, Germany\\
$^{14}$Bay Area Environmental Research Inst./NASA Ames Research Center,
Moffett Field, CA 94035\\
$^{15}$SETI Institute/NASA Ames Research Center, Moffett Field, CA 94035}
\date{Released 2011}

\def\LaTeX{L\kern-.36em\raise.3ex\hbox{a}\kern-.15em
    T\kern-.1667em\lower.7ex\hbox{E}\kern-.125emX}

\date{Accepted 2011 May 2.  Received 2011 May 2; in original form 2011 January 10}

\begin{document}

\label{firstpage}

\maketitle

\begin{abstract}
We present the results of the asteroseismic analysis of the red-giant star
KIC~4351319 (TYC 3124-914-1), observed for 30 days in short-cadence mode  
with the {\it Kepler} satellite. The analysis has allowed us to determine the large and small frequency separations,
$\Delta \nu_{0} = 24.6 \pm 0.2\, \mu$Hz and $\delta \nu_{02} =
2.2 \pm 0.3 \,\mu$Hz respectively, and the frequency of maximum oscillation power, $\nu_{\rm max} = 386.5  \pm 4.0\, \mu$Hz.\\ 
The high signal-to-noise ratio of the observations allowed us to
identify $25$ independent pulsation modes whose frequencies range approximately from $300$ to $500\,\mu$Hz.

The observed oscillation frequencies together with the accurate determination of the atmospheric parameters
(effective temperature, gravity and metallicity), provided by additional ground-based spectroscopic observations, enabled us to theoretically interpret the observed oscillation spectrum.\\ 
KIC~4351319 appears to oscillate with a well defined solar-type p-modes pattern due to radial acoustic modes and non-radial nearly pure p modes. In addition, several non-radial mixed modes have been identified.\\
Theoretical models well reproduce the observed oscillation frequencies and indicate that this star, located at the base of the ascending red-giant branch, is in the hydrogen-shell burning phase, with a mass of $\sim 1.3\, {\rm M}_{\odot}$, a radius of $\sim 3.4\,R_{\odot}$ and an age of $\sim5.6$~Gyr.
The main parameters of this star have been determined with an unprecedent level of precision for a red-giant star, with uncertainties of $2\,\%$ for mass, $7\,\%$ for age, $1\,\%$ for radius, and $4\,\%$ for luminosity.


\end{abstract}


\section{Introduction}

Oscillations excited by turbulent convection -
solar-like oscillations - have been successfully identified in cool main-sequence and post-main-sequence stars with convective envelopes. 
In the red giants, solar-like oscillations
have been first detected by ground-based observations
\citep{merline99,fra02,stello04} and by
the WIRE satellite \citep{buzasi00,retter03,retter04}.

 Difficulties in the identification of the observed modes,
led several authors \citep{guenther00,dziembowski01, teixeira03, CD04}
 to speculate on the pulsational properties of the red-giant stars.
Red giants are characterized by a deep convective envelope and a small
 degenerate helium core. Since the density in the core of these stars is quite
 large, the buoyancy frequency can reach very large values in the 
central part.
In these conditions, g modes of high frequencies can propagate and might
eventually interact with p modes giving rise to modes of mixed character.
As a consequence, the spectrum of the red giants is predicted to have a quite complicated appearance,
with a sequence of peaks uniformly spaced in frequency due to acoustic modes,
and other peaks with less clear pattern due to mixed modes.

The detection of solar-like oscillations in red giants has been well
established by the space mission MOST \citep{barban07, kallinger08a, kallinger08b}, while the CoRoT satellite was able to detect for the first time
 non-radial modes in red-giant stars
\citep{deridder09} and to find
solar-like oscillations in a very large sample of G and K giant stars \citep{hekker09, kallinger10b, mosser10} mainly lying in the core-helium-burning evolutionary phase \citep{miglio09}.  

The high-quality observations of the {\it Kepler} mission \citep{borucki10, koch10} enabled to extend the detection of
solar-like oscillations from the red clump to the lower luminosity region of the
 red-giant branch \citep{bedding10,hekker10, huber10, kallinger10a}, where stars are still burning H in the shell. This finding enables us to study
the structure and the evolution of the stars from the main-sequence phase 
up to advanced evolutionary stages by means of asteroseismology \citep{gilliland10}.

In this article we present the results of the analysis of the
oscillation spectrum of the red giant KIC~4351319 (TYC~3124-914-1) -- nicknamed `Pooh' \citep{pooh} within the
   authorship team~--
observed by the {\it Kepler} satellite for 30 days in short-cadence mode
(integration time $\sim $1 min, \citealt{gilliland10}). Here, we also consider the theoretical interpretation of the observations and
derive the asteroseismic estimates of age, mass and radius and of other
structural characteristics, making use of the accurate atmospheric parameters 
obtained by ground-based spectroscopy performed at McDonald Observatory.

 The star selected for this study is of particular interest because its
   oscillation spectrum shows an excess of power centered at frequency  $\sim380\,\mu$Hz, higher than is typical for {\it Kepler} red giants \citep{bedding10,hekker10, huber10, kallinger10a}.  This indicates that the star is well
   below the red clump in luminosity and is still ascending the red-giant
   branch \citep{miglio09}.  Its relatively high frequencies require
   {\it Kepler}'s short-cadence mode (see Section 4), which means that only relatively few stars with these characteristics are being observed by {\it Kepler}.
   KIC~4351319 therefore provides a very important opportunity to study
   this evolutionary phase and hence it has been selected for long-term observations by {\it Kepler}.

\section{Atmospheric parameters from spectroscopy}

Spectroscopic observations of KIC\,4351319 in the visible spectral range have been carried out
on 2010 July 26 (HJD\,=\,2\,455\,404.7756) with
the Cross-Dispersed Echelle Spectrograph  at the 2.7-m ``Harlan J. Smith''
telescope at the McDonald Observatory, Texas, USA, 
in the framework of the ground-based observational support of the {\it Kepler} space mission \citep{uytte10a,uytte10b}.\\
A stellar spectrum, wavelength-calibrated and normalized to the continuum, was obtained using
standard data-reduction procedures for spectroscopic observations within the NOAO /IRAF package. The
resulting signal-to-noise ratio was ${\rm SNR}\approx 100$. The spectral resolution as measured from the
Th-Ar emission lines is about R\,=\,65\,000.

In order to derive the effective temperature and surface gravity for
our target, we minimized the difference between observed and synthetic profiles. As a goodness-of-fit
test we used the parameter:
$$\chi^2 = \frac{1}{N} \sum \left(\frac{I_{\rm obs} - I_{\rm th}}{\delta I_{\rm obs}}\right)^2,$$
where $N$ is the total number of independent points in the spectrum, $I_{\rm obs}$ and $I_{\rm th}$ are the intensities
of the observed and computed profiles, respectively, and $\delta I_{\rm obs}$ is the photon noise.
The synthetic spectra were generated in three  steps: first we computed the stellar
atmosphere model by using the ATLAS9 code \citep{kur93}, then the stellar spectrum
was synthesized using SYNTHE \citep{kur81} and finally the instrumental and
rotational convolutions were applied. The ATLAS9 code includes the metal opacity by means of
distribution functions (ODF) that are tabulated for multiples of the solar metallicity
and for various microturbulence velocities.

We selected a set of
vanadium and iron lines in the range between 6180~{\AA} and 6280~{\AA}, which are free of blending
and whose atomic parameters are well known and already explored for temperature
calibrations \citep[see][and reference therein]{biazzo07}. For
surface gravity we used the non-saturated line Ca{\sc i} at\,6162.173\,{\AA}, whose wings 
are very sensitive to gravity changes \citep[see][]{grey}.

\begin{figure}
\begin{center}
\includegraphics[width=8cm]{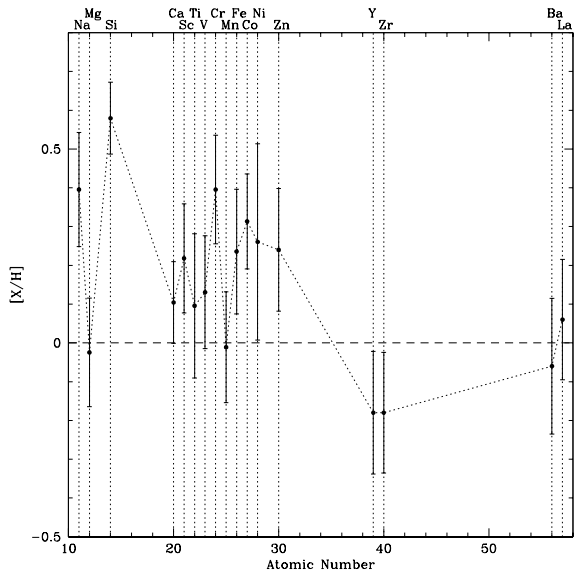}
\caption[]{Abundance pattern computed for all metals detected in the spectrum
recorded with the "Harlan J. Smith" telescope at the McDonald Observatory. The horizontal dashed line represents
the solar abundance as given in \citet{grevesse10}.}
\label{patt}
\end{center}
\end{figure}

First of all, we computed the $v\sin i$ of the star by using the parameters from the KIC catalogue \citep{latham} reported in Table~\ref{par}.
We used SYNTHE to reproduce the observed metal lines, and our best match was achieved
convolving the computed lines with a stellar rotational profile having
$v\sin i$\,=\,6\,$\pm$\,1 km s$^{-1}$, while
the microturbulent velocity $\xi=\,1\,{\rm km\,s}^{-1}$ was obtained by
using the relation $\xi\,=\,\xi(T_{\rm eff},\log g)$ published by \citet{allende04}.\\
Then, with an iterative procedure starting with the values of the KIC catalogue of $T_{\rm eff}$ 
and $\log g$ and the solar abundances given by \citet{grevesse10}, we
obtained the best fit values reported in Table~\ref{par}.
Uncertainties in  $T_{\rm eff}$ and $\log g$ have been estimated within a 1\,$\sigma$ level of
confidence, as the variation in the parameters which increases the $\chi^2$ by one
 \citep{lampton76}. The uncertainty in the iron abundance is the standard deviation of the
weighted average of the abundances derived from each single line.
\begin{table}
\begin{center}
\caption{Basic parameters of KIC~4351319}
\label{par}
\begin{tabular} {@{}lc  lc@{}}
\hline
&\multicolumn{1}{@{}l}{KIC and Simbad catalogues} & \multicolumn{2}{c}{Present analysis}\\
\hline
$V$  & 10.19 $^a$              &        &          $-$                  \\
  $ v \sin i$  & $ -$       &  & $6\pm1\,\rm{km\, s^{-1}}$ \\
$T_{\rm eff}$ & $4725\pm200\,\rm{K}$ $^b$ &  & $4700\pm50\,\rm{K}$ \\
  $\log g$ & $3.353\pm0.5$ dex $^b$  &  & $3.3\pm0.1$ dex \\
  ${\rm [Fe/H]}$ & $-0.563\pm0.5$ dex $^b$  &  & $0.23\pm0.15$ dex\\
\hline
\end{tabular}
\begin{flushleft}
$^a$ http://simbad.u-strasbg.fr/simbad/ \\
$^b$ http://archive.stsci.edu/kepler/kepler\_fov/search.php
\end{flushleft}
\end{center}
\end{table}
After having fixed $T_{\rm eff}$, $\log g$ and ${\rm [Fe/H]}$, we used SYNTHE to derive
the abundance pattern of our target by the method of spectral synthesis. The results
are summarized in Table~\ref{abundances} and plotted in Fig.~\ref{patt}.
We then conclude that, according to estimated errors, the metallicity
is slightly over the
solar ones. In particular, iron abundance is equal to
${\rm [Fe/H]}=0.23\pm0.15$\,dex.

\begin{table}
\caption[]{Summary of the inferred abundances in the atmosphere of KIC\,4351319}
\label{abundances}
\begin{center}
\begin{tabular} {lr@{$\pm$}l}
\hline
\hline
El & \multicolumn{2}{c}{ $[X/H]$}  \\ 
\hline
 Na  &    0.39&0.15\\
 Mg  & $-$0.02&0.14\\
 Si &    0.58&0.09\\
 Ca &    0.10&0.10\\
 Sc &    0.22&0.14\\
 Ti &    0.09&0.19\\
 V  &    0.13&0.14\\
 Cr &    0.39&0.14\\
 Mn & $-$0.01&0.14\\
 Fe &    0.23&0.15\\
 Co &    0.31&0.12\\
 Ni &    0.26&0.25\\
 Zn &    0.24&0.16\\
 Y  & $-$0.18&0.16\\
 Zr & $-$0.18&0.15\\
 Ba & $-$0.06&0.17\\
 La &    0.06&0.15\\

\hline
\end{tabular}
\end{center}
\end{table}

\section{Solar-like properties in red-giants stars}
 
\subsection{Large and small separations}
It is well known that the properties of solar-like oscillations in main-sequence stars 
can be described by adopting the asymptotic development \citep{tas},
which predicts that the oscillations frequencies $\nu_{n,l}$ of acoustic modes,
 characterized by radial order $n$, and harmonic degree $l$, for $l\ll n$ should satisfy the following approximation:
\begin{equation}
\nu_{n,l}\sim\Delta\nu\left(n+\frac{l}{2}+\epsilon \right),
\label{eq1}   
\end{equation}   
where $\epsilon$ is a function of frequency and depends on the properties of the surface layers and
$\Delta\nu$, known as the large frequency separation, is the inverse of the sound travel time across the stellar
diameter 
and is proportional to the square root of the mean
density.

Thus, the solar-like oscillations spectra should show a series of equally 
spaced peaks separated by $\Delta\nu$  between p modes of same degree
and adjacent $n$:
\begin{equation}
\Delta\nu\sim\Delta\nu_l=\nu_{n+1,l}-\nu_{n,l}.
\label{EQ_2}
\end{equation}

In addition,  the solar-like power spectra are characterized by another series
of peaks, whose separation is $\delta\nu_{l,l+2}$,
 known as the small separation:
\begin{equation}
\delta\nu_{l,l+2}= \nu_{n,l}-\nu_{n-1,l+2},
\label{EQ_3}
\end{equation}
which is sensitive to the chemical composition gradient in the central 
regions of the star and hence to its evolutionary state.

The determination of the large and small frequency separations
can provide asteroseismic inferences on the mass and the age of main-sequence 
and post-main-sequence solar-type stars, using 
the so-called seismic diagnostic C-D diagram \citep{CD88}.  However,
these two seismic indicators alone do not represent an unambiguous
diagnostic tool 
 for 
more evolved stars, for which the relation 
between the small and large separation
is almost linear \citep{bedding10, montalban10, mosser10}.
On the other hand, it has been demonstrated that the stellar mass and radius of a solar-like star can successfully be derived, within
  $7\,\%$ and $3\,\%$ respectively \citep[e.g.][]{kallinger10b}, 
from measurements of the large separation and $\nu_{\mathrm max}$, the frequency at which the oscillation signal reaches a maximum,
by using the scaling laws given by \citet{kjeldsen95} and \citet{bedding03}.

Hence, in order to characterize in details the structure of a red-giant star, it is necessary
to consider individual frequencies of oscillation and other oscillation properties, as we will discuss in the following.

\subsection{The potential of the mixed modes}

The properties of solar-like oscillations are expected to change
as the stellar structure evolves. 
According to Eq.~(\ref{eq1}) and considering that $\Delta\nu \propto R^{-3/2}$,
oscillation frequencies of a given harmonic degree should decrease as the star evolves and 
the radius increases   
and should be almost uniformly spaced by $\Delta \nu$ at
each stage of evolution. However, in subgiants and red giants the   
radial modes seem to follow Eq.~(\ref{eq1}) closely, but
the frequencies of some non-radial modes appear to be shifted from the 
regular spacing due to the occurrence of the so-called `avoided crossing' \citep{CD04}.
As the star evolves away from the main sequence, the core contracts 
and the radius expands, causing an increase of the local gravitational 
acceleration and of the gradients in the hydrogen abundance, and hence of
the buoyancy frequency in the deep interior of the star. 
As a consequence g modes with high frequencies
are allowed to propagate and can interact with p modes of similar frequency
and same harmonic degree, giving rise to modes with mixed character, 
which behave as g modes in the interior and p modes in the outer envelope 
\citep{aizenman77}.
The interaction can be explained as the coupling of two oscillators of similar frequencies.
The effect of 
the coupling becomes much weaker for modes with higher harmonic degree,
since in these cases the gravity waves are better trapped in the 
stellar interior and hence better separated from the region of propagation 
of the acoustic waves \citep{dziembowski01, CD04, dupret09}. 

It has been found by
\citet{montalban10} and observationally demonstrated by \citet{huber10}, that the scatter 
of $l=1$ modes caused
by `avoided crossing' decreases  as the star goes up to the red-giant branch:
as the luminosity increases and the core become denser, the $l=1$ acoustic
modes are better trapped and the oscillation spectra become more regular.
Once the star ignites He in the core, the core expands and the external 
convective zone becomes shallower which has the effect of increasing the probability of
coupling between g and p modes again.

 Very recently, \citet{beck11} have demonstrated that the quality of the {\it Kepler} observations gives the possibility to
measure the period spacings of mixed-modes with gravity-dominated character 
which, like pure gravity modes, penetrate deeply in the core allowing
 to study the density contrast between the core region and the convective envelope and, like p modes, have amplitude at the surface high enough to be observed.
In particular, 
\citet{bedding11} found that measurements of the the period spacings of the
 gravity-dominated mixed modes,
permit to distinguish between hydrogen- and helium-burning stages of evolution of the red giants.

The occurrence of mixed modes is then a strong indicator of the evolutionary state of a red-giant star and the fitting of the observed modes with those calculated by theoretical models can provide not only mass and radius but, with a good approximation, an estimate of the age of the star.

\subsection{Sharp features inside the star}

Sharp variations localized at a certain acoustic depth in the structure of pulsating stars produce a distinctive quasi-periodic signal in the  frequencies of oscillations.
The characteristics of such signal are related to the location and thermodynamic properties of the layer where the sharp variation occurs.
Sources of sharp variations are for example the borders of convection zones and regions of rapid variation in the first adiabatic exponent $\Gamma_1=
(\partial\ln p / \partial \ln \rho)_{\rm ad}$, where the derivative corresponds to an adiabatic change, such as the one that occurs in the region of the second ionization of helium.\\
In the main-sequence stars the signals
coming from different sharp features in the interior, 
overlap generating a complex behaviour \citep{mazu01}. In red giants, a recent study by \citet{miglio10} demonstrated that the oscillatory signal of the
frequencies observed is directly related only to 
the second helium ionization zone.
 
Several attempts have been made in order to isolate the generated oscillatory components from the 
frequencies of oscillations or from linear combinations of them such as
large separations and second differences.
In principle, this approach can be applied to determine the properties of the base of the convective envelope \citep{mon00, bal04} and in particular to infer the helium abundance in the stellar envelope \citep{lop97,mon98, per98, miglio03, basu04, houdek07}.



\section{{\it Kepler} observations and analysis of the oscillation spectrum}

KIC~4351319  has been observed by the {\it Kepler} satellite  \citep{koch10}
during the Q0, Q1 and Q2 runs in long cadence mode (integration time of $\sim$
 30 min)  and during the Q3 run in short cadence mode (integration time of $\sim$1 min).
The data that we describe here were obtained during 30 consecutive days,
 starting on 2009, September 18, which corresponds to the first month of observing quarter 3 (Q3.1).
The {\it Kepler} short-cadence observation mode corresponds to a Nyquist frequency of 8497 $\mu$Hz. 
In the present analysis, we used the {\it SOC corrected} data obtained
by the {\it Kepler} team after reducing and correcting the raw photometric data
 for slopes and discontinuities as described in
\citet{jenkins10} and \citet{garcia11}.
The effective duty cycle resulted to be $\sim 97\%$ and the spectral window does not show aliasing
artifacts that could interfere with the identification of real frequencies. 

\begin{figure*}
\centering
\includegraphics[draft=false, scale=0.7]{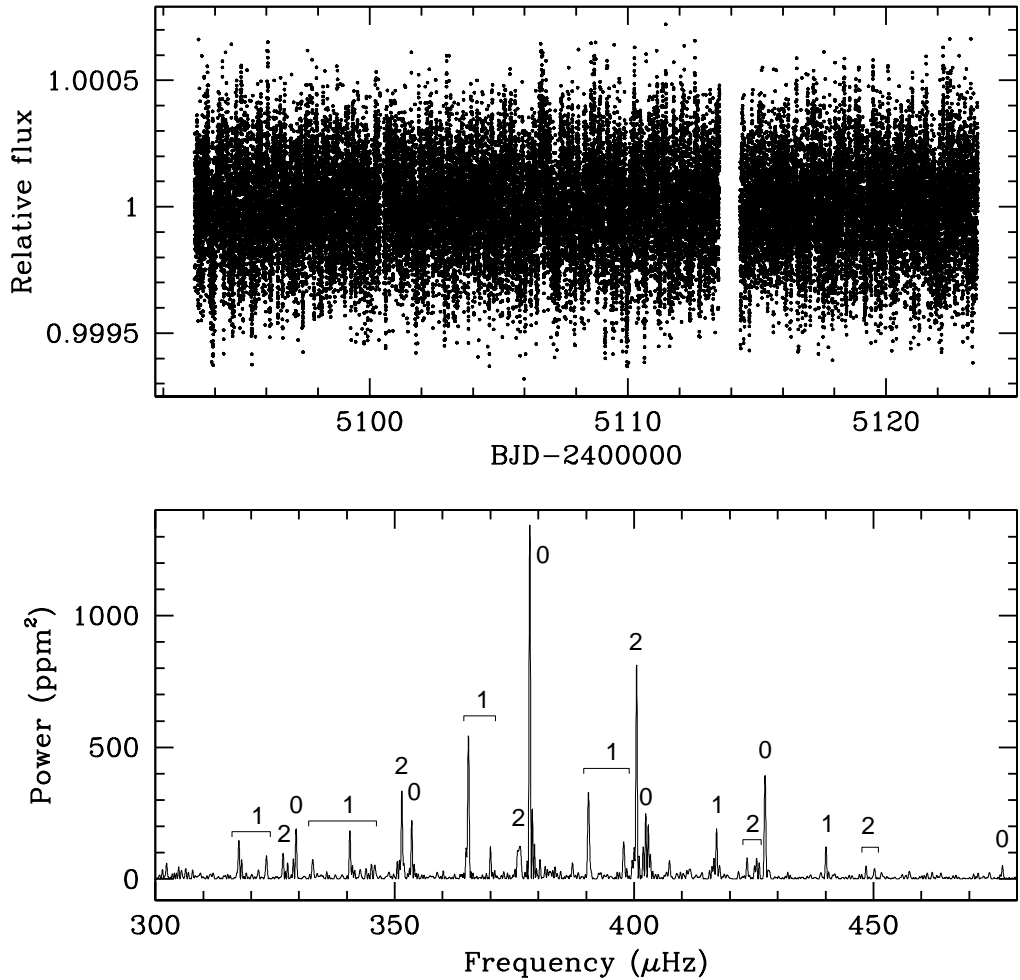}
\caption{Top panel:Light curve corrected for slopes and discontinuities and normalized to the mean value. Bottom panel: Fourier spectrum in the region of
p mode excess bump; harmonic degree of the excited modes are also shown}.
\protect\label{curvaluce}
\end{figure*}

The top panel of Fig.~\ref{curvaluce} shows the cleaned light curve normalized to the mean value.

The Fourier analysis of the data set has been performed by using the PERIOD04 package \citep{lenz},
which allows to extract the individual frequencies from large  multiperiodic time series - also containing gaps -
and can perform multiple-frequency least-squares fit up to several hundred simultaneous sinusoidal components. 
The resulting power spectrum of KIC 4351319 shows a clear power excess in the frequency range $300 - 500\, \mu$Hz with regularly spaced peaks (bottom panel of Fig.~\ref{curvaluce}), while no
relevant features appear between $550\, \mu$Hz and the Nyquist limit.
We have considered only peaks in the spectrum with a signal-to-noise ratio (SNR) in amplitude greater than 3.


\begin{table}
\begin{center}
\caption{KIC 4351319: observed oscillation frequencies with errors, their SNR and harmonic degree. The symbol $\bullet$ denotes modes confirmed by all the teams.}
\label{tab_final}
\begin{tabular}{lr@{.}ll}
\hline\hline
 Freq ($\mu$Hz)& \multicolumn{2}{c}{SNR} &{$l$}\\
\hline
  329.16 $\pm 0.24$            &   5&4  & 0   \\
  353.62 $\pm 0.20$ $\bullet$  &   5&4  & 0   \\
  378.34 $\pm 0.11$ $\bullet$  &  13&1  & 0   \\
  402.43 $\pm 0.07$ $\bullet$  &   5&5  & 0   \\
  427.32 $\pm 0.12$ $\bullet$  &   7&5  & 0   \\
  476.86 $\pm 0.54$            &   3&6  & 0   \\
  317.38 $\pm 0.19$ $\bullet$  &   4&1  & 1     \\
  323.00 $\pm 0.20$            &   3&4  & 1     \\
  332.98 $\pm 0.30$            &   3&2  & 1     \\
  340.67 $\pm 0.09$ $\bullet$  &   5&6  & 1     \\
  345.26 $\pm 0.66$            &   3&2  & 1     \\
  365.40 $\pm 0.07$ $\bullet$  &   8&9  & 1    \\
  370.09 $\pm 0.15$            &   3&5  & 1     \\
  390.50 $\pm 0.09$ $\bullet$  &   6&1  & 1     \\
  397.89 $\pm 0.20$            &   3&8  & 1     \\
  417.30 $\pm 0.11$ $\bullet$  &   5&1  & 1     \\
  440.20 $\pm 0.10$ $\bullet$  &   4&7  & 1     \\
  326.55 $\pm 0.25$            &   4&0  & 2     \\
  351.32 $\pm 0.15$ $\bullet$  &   7&2  & 2     \\
  376.03 $\pm 0.19$ $\bullet$  &   5&0  & 2     \\
  400.56 $\pm 0.08$ $\bullet$  &   9&4  & 2     \\
  423.71 $\pm 0.40$            &   3&4  & 2     \\
  425.80 $\pm 0.32$ $\bullet$  &   3&2  & 2     \\
  448.65 $\pm 0.40$            &   3&4  & 2     \\
  449.98 $\pm 0.50$            &   3&1  & 2     \\
\hline
\end{tabular}
\end {center}
\end{table}

In order to confirm the results obtained,
the analysis of the spectrum was also independently
performed
 by five additional teams which
 adopted the following
 methods:
\begin{enumerate}
\item[a)]
 the power spectrum was smoothed
 with a Gaussian with a FWHM of 0.5 $\, \mu$Hz and the frequencies of the peaks with $\rm{SNR} > 3$ (in amplitude) were
measured.
\item[b)]
 peak-bagging methods, substantially based on the fit of Lorentzian profiles to the power density spectrum, using:
\begin{itemize}
\item Maximum Likelihood Estimators (MLE), (see Appendix~A for details)
\item Bayesian MCMC \citep{Gruberbauer09}, (see Appendix~B for details)
\item Envelope autocorrelation function analysis \citep{mosser09}
\end{itemize}
\end{enumerate}

Table~\ref{tab_final} reports the set of frequencies
confirmed, within the errors, by
at least two of the peak-bagging solutions.
Frequencies  labeled with ($\bullet$) have been detected
by all the teams.

The values of the radial order for $l=0$ are $n=12-18$ as obtained by
adopting the scaling law 
proposed by \citet{mosser11} \citep[see also][]{huber10}.

Three possible excited modes with $2\leq \rm{SNR} \leq 3$, namely
 $302.32\,\mu$Hz,
$ 387.12\, \mu$Hz and $407.41\, \mu$Hz have been also found,
but their validity must be carefully checked by additional future observations.



\begin{figure}
\centering
\includegraphics[draft=false,scale=0.45]{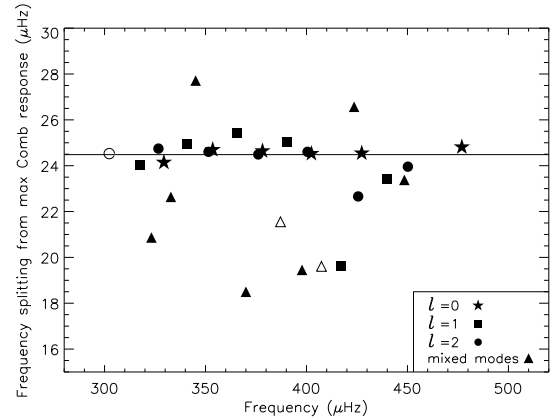}
\caption{Large frequency separation as
determined from the comb-response analysis
applied to the frequencies of Table~\ref{tab_final} (filled symbols) and to three possible frequencies with $2\leq \rm{SNR} \leq 3$ (open symbols). The horizontal line shows the
mean value of $24.6 \pm 0.2\, \mu$Hz computed for the $l=0$ modes of Table~\ref{tab_final} only.
The reported uncertainty is the standard deviation of the mean. Different symbols refer to $l=0$ (stars), $l=1$ (squares)
and $l=2$ (circles) modes, respectively, as derived by the alignment in the \'echelle diagram. Triangles refer to mixed modes (see text for more details).}
\protect\label{comb}
\end{figure}

In order to estimate the large frequency separation, 
we first computed the autocorrelation function of the power spectrum in the region of p-mode power excess.
 With such an initial
 guess value of the large separation,
 we built an \'echelle diagram and identified the $l=0$ ridge.
Afterwards, we computed the comb-response function,
 as defined by \citet{kjeldsen95b},
for each frequency reported in Table~\ref{tab_final} and for the three additional frequencies with $2\leq \rm{SNR} \leq 3$, searching for the maximum value of
the comb response in the range $(
15 - 30)\,\mu$Hz. The result,
 reported in Fig.~\ref{comb}, shows that many
 modes lie around the mean value of $\langle\Delta\nu_0\rangle = 24.6 \pm 0.2\, \mu$Hz computed for the radial modes, while 
the other frequencies
shows a quite large scatter indicative of 
their
 mixed mode character.

\begin{figure}
\centering
\includegraphics[draft=false,scale=0.45]{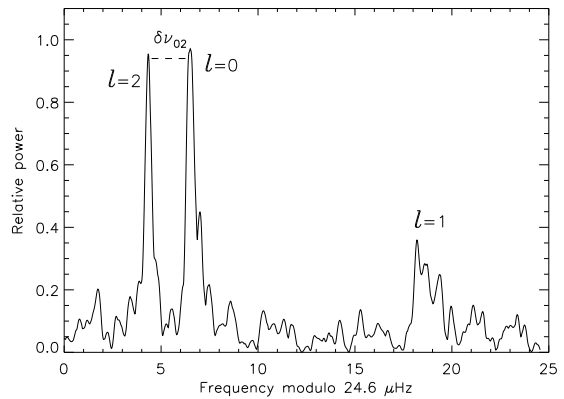}
\caption{The power
spectrum of KIC 4351319 folded at the average large separation, showing clear peaks corresponding to $l=0, 1$ and $2$. The small frequency separation $\delta \nu_{02}$ is also shown.}
\protect\label{power2}
\end{figure}

Fig.~\ref{power2} shows the region from 250 to 650 $\mu$Hz of the power spectrum folded with a spacing equal to
the mean value of the large frequency separation. Two sharp peaks, corresponding to modes with $l=0$ and $l=2$, as well as a broader peak, corresponding to modes with $l=1$, can be clearly seen. The final mode identification in terms of the degree $l$, based on both the alignment in the \'echelle diagram (see Fig. \ref{echelle}) and the result of the peak bagging  is reported in Table~\ref{tab_final}. A value of the mean small frequency separation $\langle\delta\nu_{02}\rangle = 2.2 \pm 0.3\, \mu$Hz has been also derived from each pairs of $\nu_{n,0}$ and  $\nu_{n-1,2}$ reported in Table~\ref{tab_final}.


\subsection{Background and p-mode power excess modelling}

After converting the power spectrum (ppm$^2$) to power spectral density
(ppm$^2 \mu \rm{Hz}^{-1}$), by multiplying the power by the effective length of the observing run --
estimated as the reciprocal of the area under the spectral window \citep{bedding05}  -- a model fitting of the stellar background due to stellar activity and granulation has been performed.
In order to simultaneously model the stellar background as well as the p-mode power excess, we fitted the observed power density spectrum by a superposition of white noise, two semi-Lorentzian functions \citep{harvey85} and a
Gaussian function representing the power excess hump \citep[see][for details]{kallinger10b}:

\begin{equation}
 P(\nu) = \sum_{i=1}^{2} \frac{A_i}{(1+\nu B_i)^{c_i}} + D e^{- \frac{(\nu_{\mathrm max} -\nu)^2}{2 \sigma^2}} + E
\label{EQ_4}
\end{equation}
where $\nu$ is the frequency, $A_i$, $B_i$ and $c_i$ are the amplitudes, the characteristic timescales and the slopes of the power laws; $E$
 is the white noise contribution; $D$, $\nu_{\mathrm max}$ and $\sigma$ are the height, the central frequency and the width of the power excess hump, respectively.

\begin{figure*}
\begin{center}
\includegraphics[draft=false,scale=0.9]{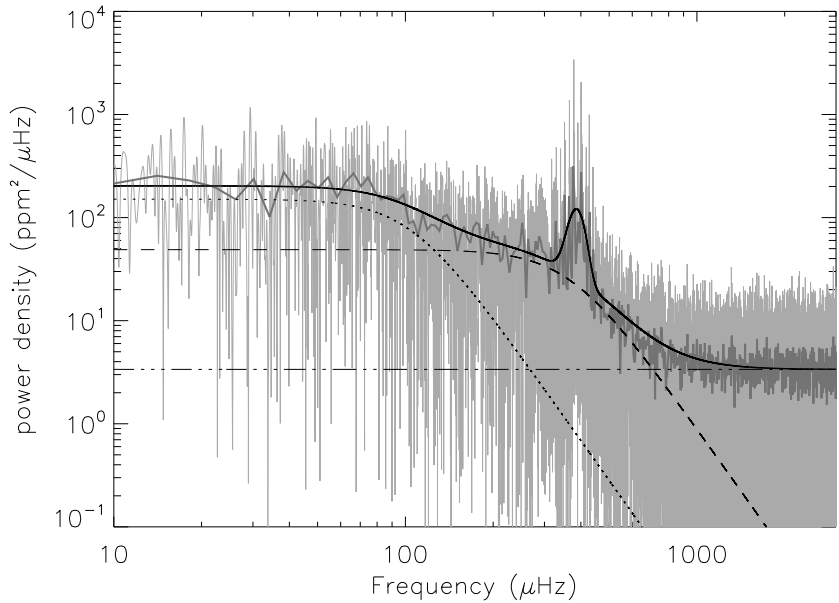}
\caption{Power density spectrum of KIC 4351319. Light grey: power density spectrum; dark grey:
the raw spectrum smoothed with a $4\,\mu$Hz boxcar taking only the independent averages; thick solid line: global fit according to the model described in the text; dashed, dotted and dashed-dotted lines: the two semi-Loretzians and the white noise component of the global fit, respectively.}
\label{back}
\end{center}
\end{figure*}

We adopted as initial guess for the white noise the average of the power spectrum in the frequency range
$1000 - 2000\,\mu$Hz, where the photon noise is expected to dominate over the other components. The guess
values for the remaining parameters have been evaluated by scaling from typical solar values \citep{aigrain04,
palle99, huber09} for active region and granulation timescales. We smoothed the raw power spectrum density by applying a $4\,\mu$Hz boxcar using the independent averages only, as suggested by \citet{garcia09}, in order to perform properly the fit by means of a weighted least-squares procedure. The weights 
we adopted, 
as in \citet{garcia09}, have been determined by the uncertainties on each independent $4\,\mu$Hz average.
As in \citet{mathur10} we adopted the IDL least-squares MPFIT package (provided by Craig B. Markwardt, NASA/GSFC\footnote{ http://cow.physics.wisc.edu/$\sim$craigm/idl/idl.html}), implementation of the Levenberg-Marquardt fitting algorithm.  The results are shown in Fig.~\ref{back}.

The fit returned, 
among other parameters, the maximum amplitude of the smoothed excess power P$_{\rm max} = 94.99 \pm 0.3\,{\rm ppm}^2/\mu$Hz. 
The corresponding frequency is
 $\nu_{\rm max} = 386.5 \pm 4.0\,\mu$Hz, 
where the uncertainty is given by the resolution of the smoothed spectrum.
 This value and its uncertainty are in a very good agreement with what found by fitting the raw spectrum by the 
envelope autocorrelation function analysis \citep{mosser09}.\\
Following \citet{kjeldsen08}, we evaluated the maximum amplitude
per radial mode by adopting the calibration factor c = 3.16, whose value
depends on the spatial response of the observations to modes with
different degree relative to radial modes, and obtained ${\rm 
A_{\mathrm max}}
\simeq(
27.19\pm 0.16)$ppm.

\subsection{Global asteroseismic parameters}

In order to extract a rough estimate of the global parameters of the star, to be adopted as guess values for
the model computation, we adopted the scaling laws provided by \citet{kjeldsen95} and \citet{bedding03}, relating the observed mean large frequency separation $24.6 \pm 0.2\, \mu$Hz and $\nu_{\rm max} = 386.5 \pm 4.0\,\mu$Hz to the fundamental parameters of the star.
We obtained: $M/{\rm M}_{\odot}\simeq 1.35\pm 0.09$ and $ R/{\rm R}_{\odot}\simeq 3.44\pm 0.08$. We also derived the expected radial order of the mode with maximum amplitude in the spectrum, $n_{\rm max}\simeq 14$.
This latter value is in agreement
 with the radial order identification of the radial mode with higher SNR reported in Sect.~4.

\section{On the characterization of the structure of KIC~4351319}

\subsection{Evolutionary models}
\begin{figure}
\centering
\includegraphics[draft=false,scale=0.45]{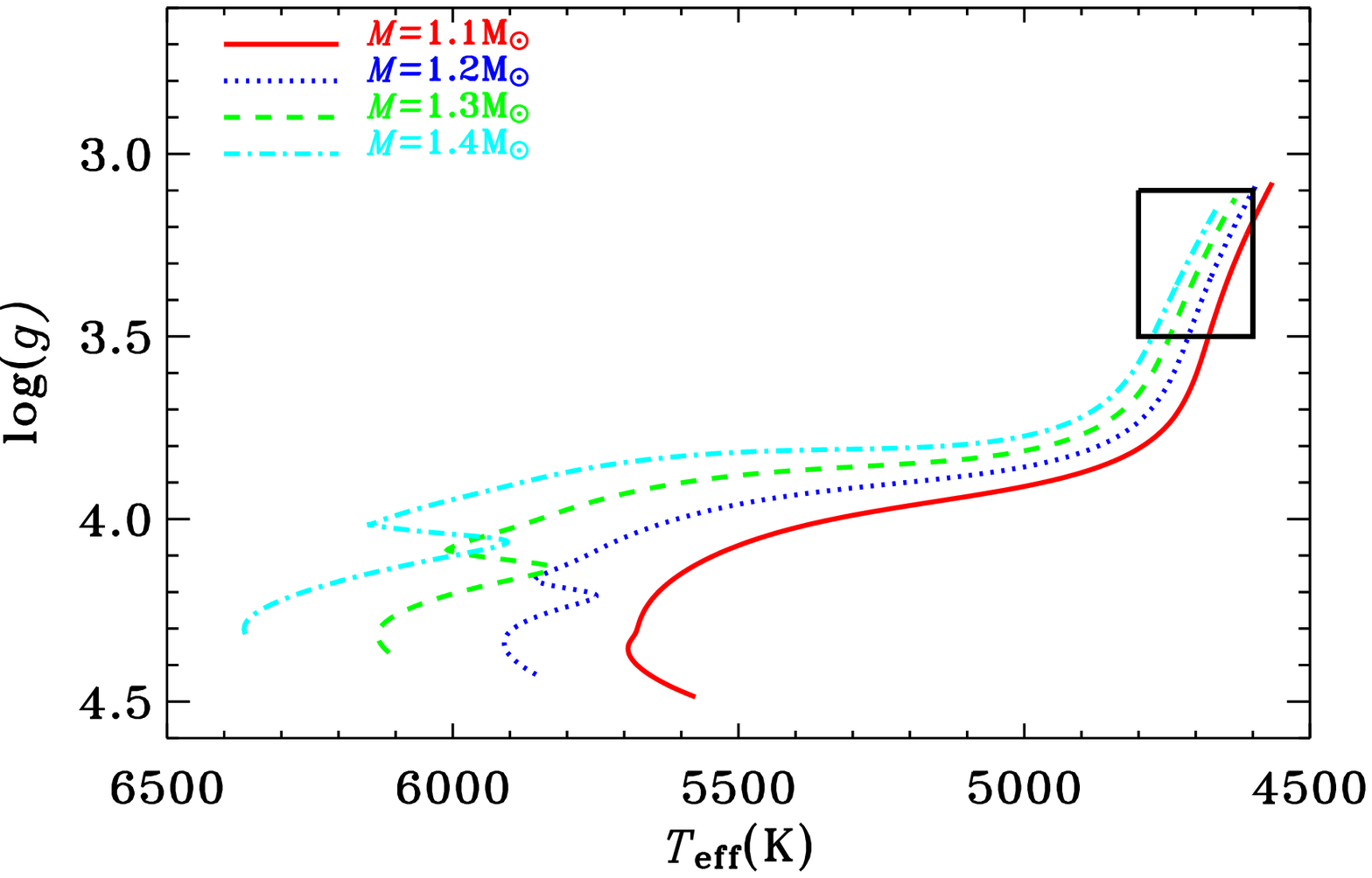}
\includegraphics[draft=false,scale=0.45]{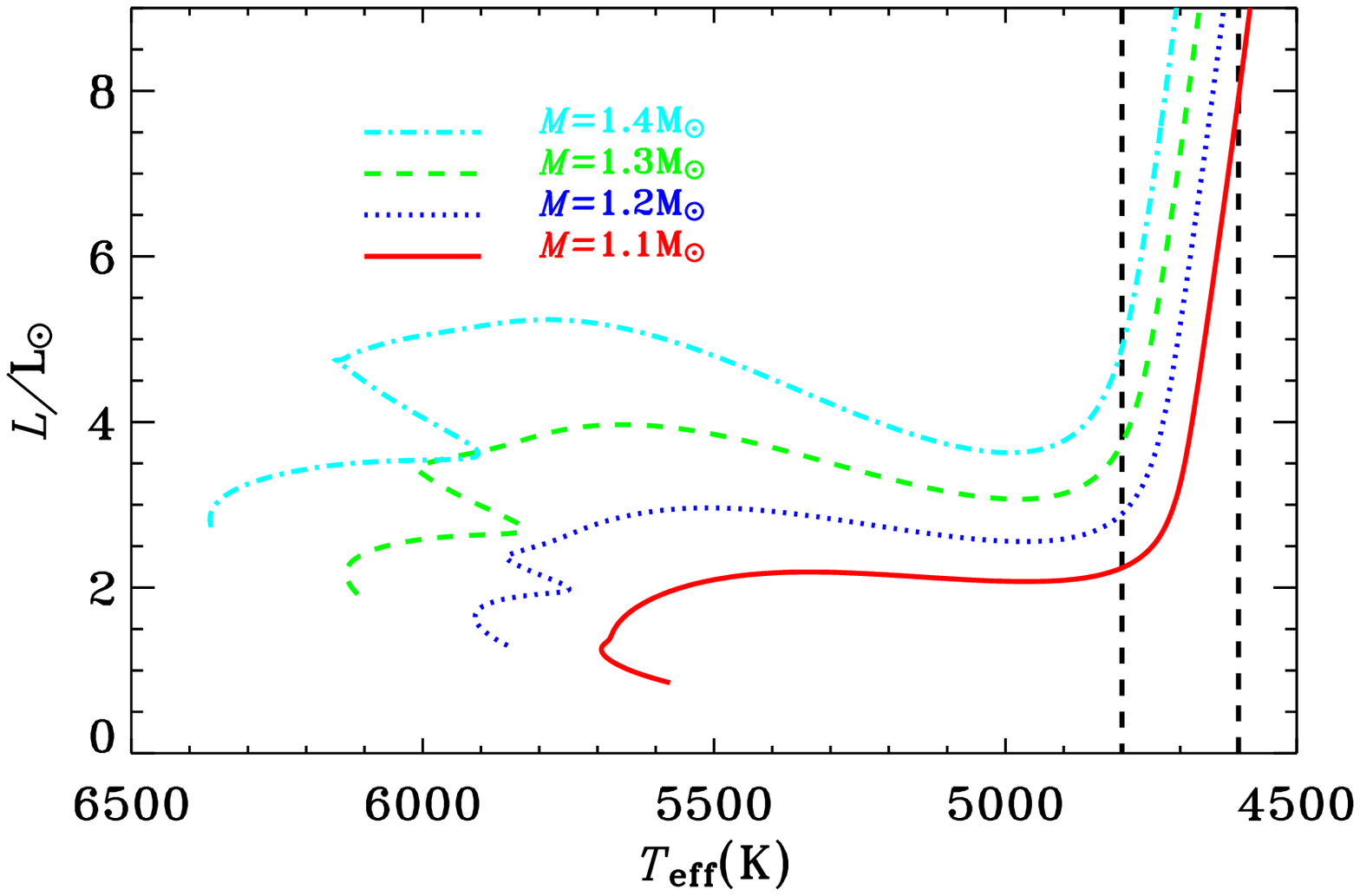}
\caption{Evolutionary tracks plotted in the H-R diagrams (the upper 
panel shows $T_{\rm eff} - \log g$, the lower panel shows $T_{\rm eff} - L$)
 calculated for increasing values of the mass, while all the other parameters are fixed. The metallicity is $Z=0.03$, the initial hydrogen abundance $X_0=0.7$, and $\alpha=1.8$. Here, no additional effects are included.  
 The rectangle defines the two-sigma error box for the observed gravity and effective temperature. In the bottom panel, black dashed lines indicate the uncertainty in the observed effective temperature while observed values of luminosity are not known. } 
\protect\label{HR}
\end{figure}

Given the identified pulsation frequencies and the basic
photospheric parameters, we faced the theoretical challenge
to interpret the observed oscillation modes by constructing stellar models which 
satisfy the observational constraints.

We assumed the effective temperature and gravity
as calculated in Section 2
respectively, $T_{\rm eff}=4700\pm50$K, $\log g=3.3\pm0.1$~dex (Table \ref{par}). 

We calculated several grids of theoretical structure models for the star by using the ASTEC evolution
code \citep{CD08a} by varying the mass and the composition in order to match the atmospheric parameters available. 

All the models have been calculated with the OPAL 2005 equation of state 
\citep{OPAL}, OPAL opacities 
\citep{Igl96}, and the NACRE nuclear reaction rates \citep{NACRE}. Convection was treated according to the
mixing-length formalism (MLT) \citep{bohm} and defined through the parameter $\alpha=\ell/H_p$, where $H_p$ is the pressure scale height and $\alpha$ is assumed to vary from $1.6$ to $1.8$. 

Inclusion of overshooting or diffusion outside
 the convective core during the main-sequence phase did not produce any appreciable effect on the oscillation frequencies of models in such evolved phases of the evolution. 

A crucial input quantity is the iron abundance,   
whose logarithmic value relative to the solar one has been taken as 
[Fe/H]$=(0.23\pm0.15)$~dex, as determined in Sect. 2. 
The initial heavy element mass fraction $Z$
can be calculated from the relation [Fe/H]$=\log(Z/X)-\log(Z/X)_{\odot}$, where $(Z/X)$ is the ratio at
the stellar surface and the solar value is $(Z/X)_{\odot}=0.0245$ \citep{GN93}. Thus, it has been assumed $Z/X=0.04\pm0.01$. 
In the present modelling we neglect the difference in the relative abundances of
the heavy elements between the star and the Sun (cf. Fig. 1); although this should be
taken into account in future, more detailed investigations, the effects on the opacity
and hence the model structure are likely to be contained within the assumed range of
uncertainty in $Z$.
The uncertainty in the observed value of $Z$ introduces an uncertainty in the determination of the mass   
whose value, considering only the observed spectroscopic parameters, seems to be limited to the range $M=(0.8-1.5)\,{\rm M}_{\odot}$.   
   
Extra mixing outside the convective region was obtained by adding
turbulent diffusion from the convective envelope, as described in 
\citet{proffitt}, by using the parameterized formulation:
\begin{equation}
 D=D_{\rm max}\left(\rho/\rho_{\rm b_{cz}}\right)^{-3}
\end{equation}
where 
 $\rho_{\rm b_{cz}}$ is the density at the base of the convective envelope and
$D_{\rm max}$ is a parameter which sets the maximum diffusion.

Additional evolutionary models were calculated by including overshoot beneath
the convective envelope by a distance  
$\ell_{\rm ov}=\alpha_{\rm ov} H_{p, {\rm b_{cz}}}$, 
where  $ H_{p,{\rm b_{cz}}}$ is the pressure scale height at the base of the convective envelope and $\alpha_{\rm ov}$ is 
a nondimensional parameter. 

The resulting evolutionary tracks are characterized by
the mass $M$, the
initial chemical composition and a
 mixing-length parameter.  Fig.~\ref{HR} shows a series of evolutionary tracks
 obtained for different masses and fixed initial composition and
plotted in two H-R diagrams, representing respectively the effective temperature-gravity plane and the effective temperature-luminosity plane.
   
The location of the star in the Hertzsprung-Russell diagram    
identifies KIC 4351319 as being in the post-main-sequence phase of evolution at the beginning of the ascending red-giant branch.  
It has a small, degenerate helium core, having exhausted its central hydrogen,   
and it is in the shell-hydrogen-burning phase, with a very deep convective zone,
extending from the base located at    
about $r_{\rm b_{cz}}\simeq0.2R$ to the photosphere. 

\begin{figure}
\centering
\includegraphics[draft=false,scale=0.45]{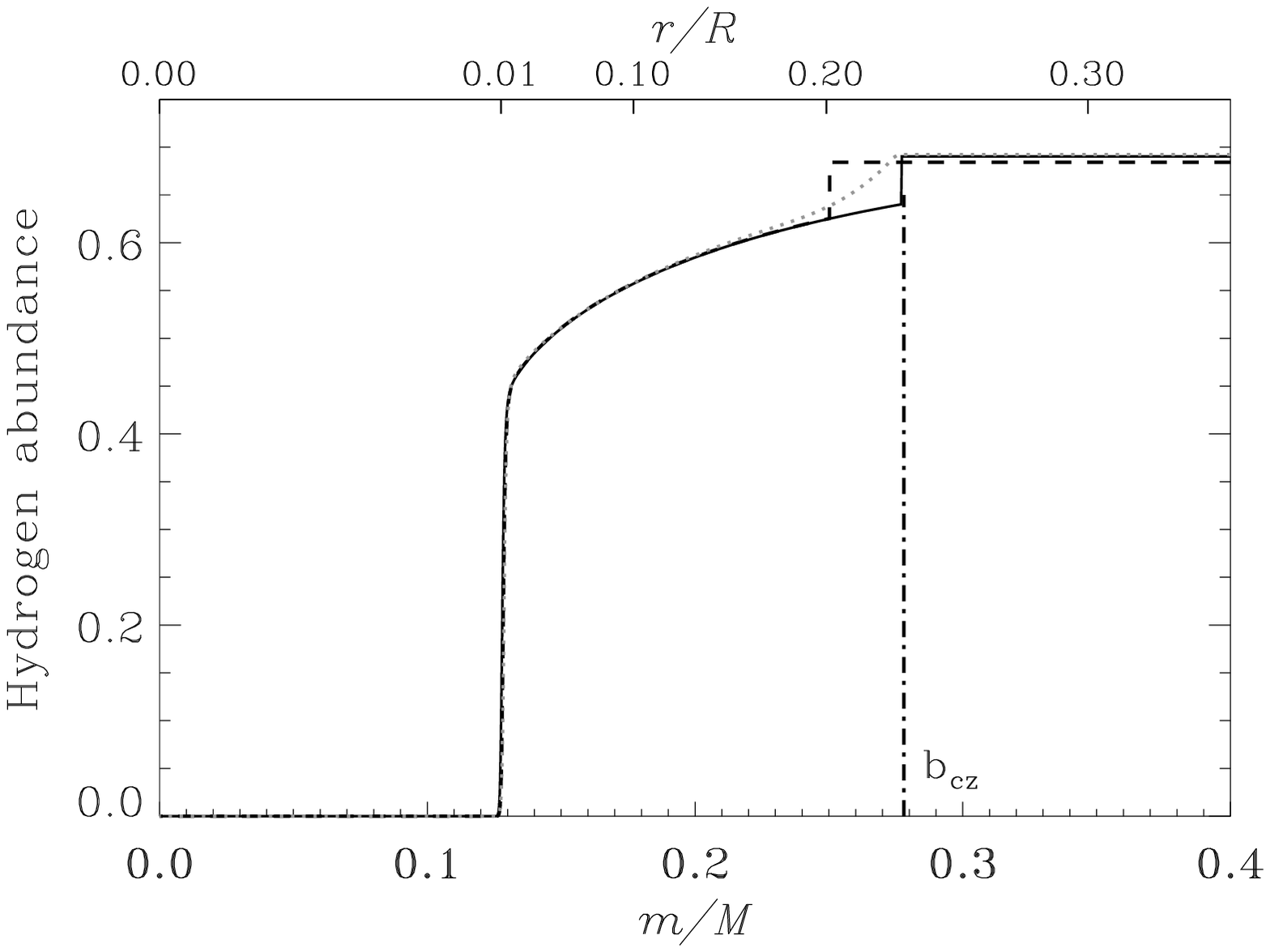}
\caption{Hydrogen content in three models of KIC~4351319 computed with $M=1.32{\rm M}_{\odot}$, $X_0=0.7$, $Z_0=0.03$, $\alpha=1.8$. The base of the convective envelope located at $m_{\rm b_{cz}}=0.28M$ and ${\rm r}_{\rm b_{cz}}=0.23R$, is shown by the dot-dashed line. The solid line corresponds to Model~1 in Table \ref{T.6} with no additional effects. Dotted line corresponds to Model~3 calculated with inclusion of strong turbulent diffusion ($D_{\rm max}=40000\,{\rm cm}^2\,{\rm s}^{-1}$).
Dashed line corresponds to Model~4 of Table~\ref{T.6} calculated with overshooting from the convective envelope.}
\protect\label{H}
\end{figure}

Figure~\ref{H}
 shows the internal content of hydrogen in three models of this star, which  
include respectively extra mixing by turbulent diffusion from the convective envelope,  overshooting from the convective envelope with $\alpha_{ov}=0.2$ and without any extra mixing effect.
 The internal hydrogen profiles
show that the core, where the hydrogen is exhausted, is very small and
a sharp variation, as a step-like function,
 occurs at the base of the convective envelope located at $m_{\rm b_{cz}}=0.28M$ and
 $r_{\rm b_{cz}}=0.23 R$ for the model with no extra effects
 considered. 
Models with inclusion of turbulent diffusion show a smooth profile of the abundance of the hydrogen with no sharp variation at the base of the envelope. 

We expect to be able to distinguish among the different internal characteristics of the structure by studying the oscillation properties  of this star.
Furthermore, we wish to verify whether, as demonstrated by \citet{CD93} for the case of the Sun, the inclusion of diffusion in the models results in a better agreement between theory and observations.

\subsection{The seismic properties of the models}

\begin{figure*}   
\centering
\includegraphics[draft=false,scale=0.7]{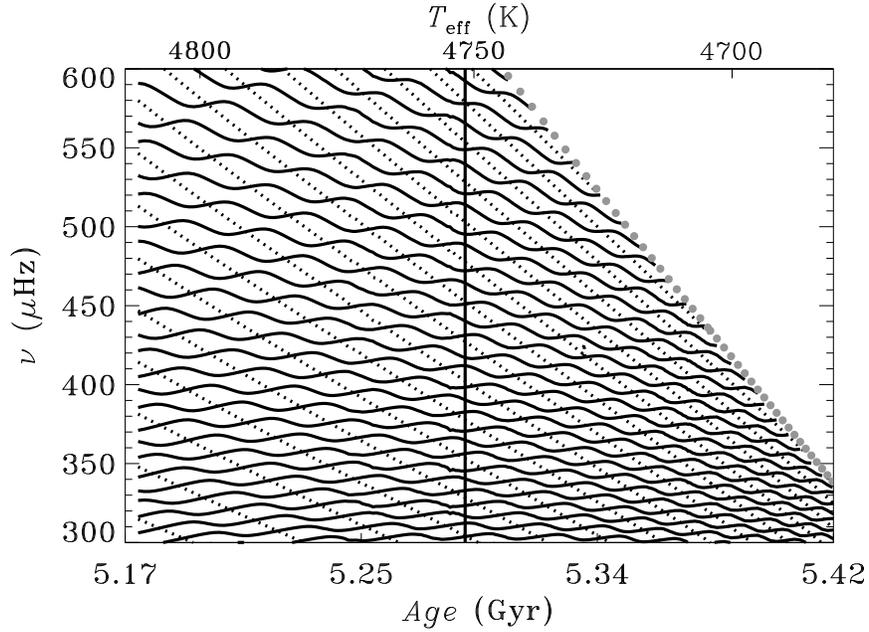}
\caption{Evolution of adiabatic frequencies with age and with effective temperature of a model computed with a mass
$M=1.32{\rm M}_{\odot}$ and $Z=0.03$. The dotted lines correspond to modes of    
degree $l=0$, and the solid lines to modes with $l=1$.   
The frequency and effective temperature ranges correspond to the observed ones. The grey dots indicate the acoustical cutoff frequency. The vertical line indicates Model~1 of Table~\ref{T.6}.}
\label{F.8}
\end{figure*}   

\begin{figure*}   
\centering   
\includegraphics[draft=false,scale=0.7]{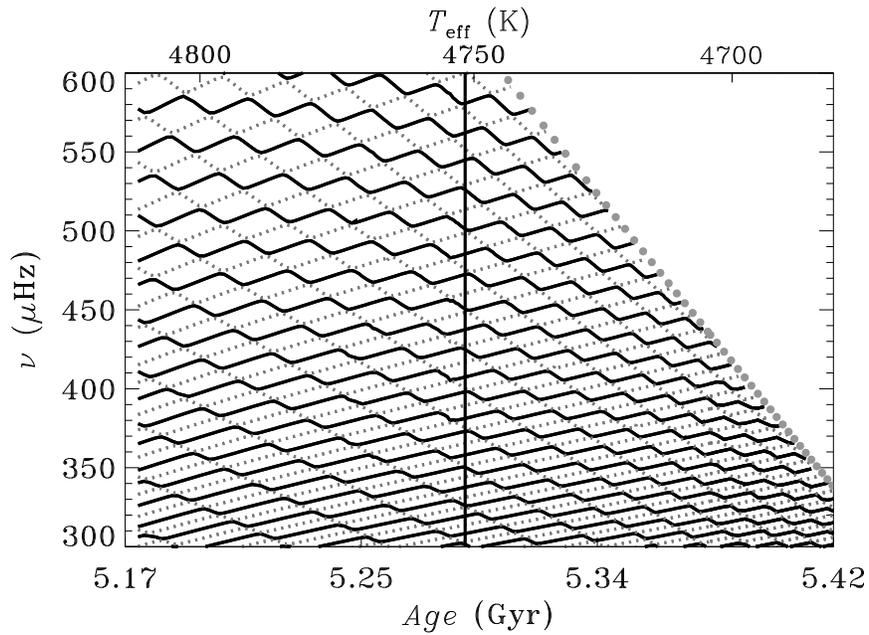}   
\caption{
Evolution of adiabatic frequencies for modes with $l=2$ as function of the age and the effective temperature for a model of KIC~4351319 computed with a mass
$M=1.32{\rm M}_{\odot}$ and $Z=0.03$. Here solid lines correspond to modes with even radial order, grey dotted lines correspond to modes with odd radial order.   
The frequency and effective temperature ranges correspond to the observed ones. The vertical line indicates Model~1 of Table~\ref{T.6}.}
\label{F.10}   
\end{figure*}

In order to investigate the observed solar-like oscillations, we used the
ADIPLS package \citep{CD08b} to compute
adiabatic oscillation frequencies with degree $l=0,1,2,3$ for all the models
satisfying the spectroscopic constraints.

Figure~\ref{F.8} shows   
the evolution of frequencies computed for an evolutionary model of KIC~4351319 calculated   
with $M=1.32{\rm M}_{\odot}$,    
$Z=0.03$,    
without additional extra mixing effects.
The ranges in frequency and effective temperature have been chosen to correspond approximately to the observed ranges.     The location of the acoustical cut-off frequency, decreasing with   
increasing age, at the top of the atmosphere in the 
model has been   
indicated by grey dots. 

  As it has been already explained in Sect.~3, 
according to Eq. (\ref{eq1}), the plot should be characterized   
by frequencies which decrease as the star evolves    
and almost uniformly spaced by $\Delta \nu$ at   
each stage of evolution.    
In the case of more evolved stars, as it has been shown by \citet{dim03} and 
\citet{metcalfe10},
 while the   
radial modes seem to follow closely Eq. (\ref{eq1}),   
the frequencies of the modes with $l=1$ show the typical oscillating behaviour
 due to the presence of
the mixed modes.
In addition, Fig. \ref{F.8} shows that at each stage of the evolution the distance between two frequencies of adjacent orders become smaller and smaller as the frequency decreases:
 while the upper part of the panel hosts mixed modes with prevalent p-mode character, below $350 \,\mu$Hz the panel appears more crowded
by mixed modes with predominant g-mode character and the general trend of frequencies decreasing as the star evolves does no longer hold.\\
In particular,  pure g modes of high radial order are located at very low frequencies and high radial orders, below $300 \,\mu$Hz.
 
Fig. \ref{F.10} shows the evolution of the frequencies for $l=2$ modes as function of the age and the effective temperature. 
It shows that $l=2$ modes, similarly
 to what happen for $l=1$ modes,
undergo avoided crossings during the evolution and the theoretical oscillation spectrum is as a result very crowded.
 The presence of mixed modes with prevalent g mode character is also evident for $l=2$: below $400 \,\mu$Hz frequencies do not appear to decrease with the decrease of the effective temperature. 

\begin{figure}   
\centering  
\includegraphics[draft=false,scale=0.45]{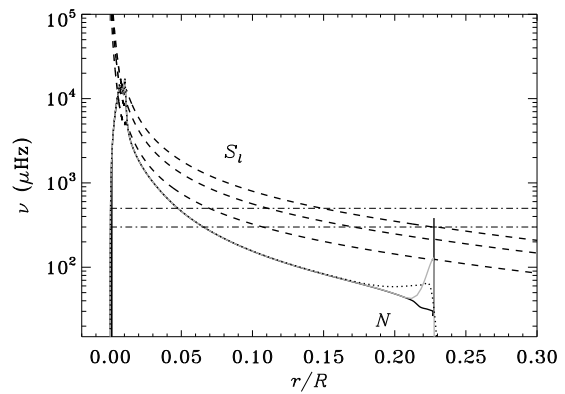}
\caption{Propagation diagram from the centre to $ r=0.3\,R$ for three models.
  The solid black line represents the buoyancy frequency $N$ for the Model 1 in Table \ref{T.6} without extra-mixing effects, the solid grey line represents the buoyancy frequency $N$ for the Model 2 in Table \ref{T.6}
with the addition of a weak diffusion from the convective 
envelope, the dotted line represents the buoyancy frequency $N$ for the Model 3 in Table \ref{T.6} with an efficient turbulent diffusion from the convective 
envelope.
 The dashed lines represent the Lamb frequencies $S_{l}$ for $l=1,2,3$.    
 The dot-dashed horizontal lines define the limits of the range of the observed frequencies.}
\protect\label{F.6}   
\end{figure}

 The behaviour of the mixed modes inside the star  
can be understood by considering the propagation diagram in   
Fig. \ref{F.6} obtained for three models of KIC~4351319, calculated respectively without extra mixing effect and with the inclusion of a weak and an efficient turbulent 
diffusion below the convective envelope.   The characteristics of the non-radial modes depend on the separation between gravity and acoustic domains, defined respectively by the 
Brunt-V\"ais\"al\"a ($N$) and Lamb frequencies ($S_l$). 
 The Lamb frequencies appear close to the g-modes region. Thus,
the gravity waves, trapped in a region not so well   
 separated from the region of propagation    
of the acoustic waves, can interact with the p modes of similar frequencies.
In addition, it is possible to notice that
at nearly the same value of effective temperature and luminosity,   
the model without extra-mixing effect shows a second maximum in    
its buoyancy frequency located at the base of the convective envelope ($r=0.23R$). This maximum becomes smaller and totally disappears as the effect of
the helium diffusion beneath the base of the convective envelope becomes stronger.
The characteristics of this second maximum in $N$ reflects the behaviour of the hydrogen abundance inside the star as seen in Fig.~\ref{H} and it will be discussed in Sect.~5.4.

 It is important to notice that, according to the theory, the spectra of g modes are denser than that of p modes: many g modes can interact with the same p mode of close frequency.

\begin{figure} 
\centering  
\includegraphics[draft=false,scale=0.45]{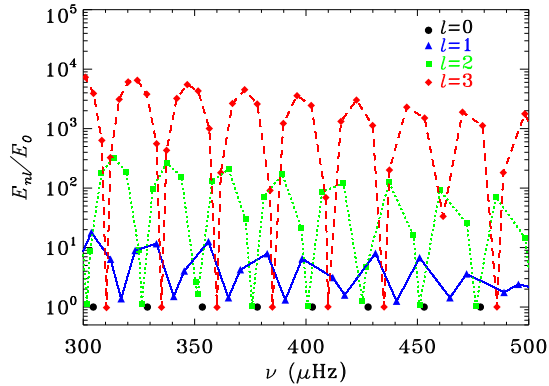}
\caption{Normalized inertia plotted against frequency for  Model~1 described in Table \ref{T.6}. Black circles are employed for $l=0$ modes. Blue triangles connected by solid lines are employed for $l=1$ modes. Green squares connected by dotted lines are employed for $l=2$ modes. Red diamonds connected by dashed lines indicate $l=3$ modes.}
\protect\label{F.12}   
\end{figure}

In order to have an idea about the modes expected to be visible,
amplitudes of all the calculated modes were roughly estimated by making the following considerations.
The total energy of a mode of radial order $n$ and harmonic degree $l$ can be expressed as ${\cal E}_{nl}=A{^2}_{nl}{E}_{nl}$, where $A_{nl}$ is the surface amplitude, and ${E}_{nl}$ is the inertia of the mode.
For stochastically excited modes it is generally expected that
the mode energy is independent of the harmonic degree at fixed frequency.
It follows that the amplitude of a mode can be estimated relative to a radial mode of the
  same frequency, such that the modes would have the same total energy
  \citep[e.g.][]{CD95, dupret09}:
\begin{equation}
\frac{A_{nl}}{A_0}=\left[\frac{E_0}{E_{nl}}\right]^{1/2},
\label{ampl}
\end{equation}
where $A_0$ and ${E}_0$ have been obtained by interpolating to the frequencies of radial modes.
Figure \ref{F.12} shows the normalized inertia for the frequencies of  Model~1 of Table~\ref{T.6} of KIC~4351319. It shows that radial modes have a very low inertia relative to most of the nonradial modes and are the
modes which will show highest amplitude at the surface. 
Some modes with $l=1$, $l=2$ and $l=3$ have an inertia similar to that of the radial modes $E_{nl}\simeq E_0$, and appear quite regularly spaced in frequency.
These modes which preserve strong p-mode 
character are solar-like oscillations and are predicted to be visible at the surface with a quite high amplitude.\\
The effect of    
 coupling, typically expected only for $l=1$, strongly appear also for modes with $l\geq 2$. 
As the inertia grows, during the coupling the gravity character of the modes predominates on the acoustic character.
Modes with the highest inertia are evanescent towards the surface and hence have the lowest probability to be observed. 
Thus, due to the low amplitude, the probability to detect any mixed modes for $l=3$ appears quite low, but it is quite likely that we can observe many of the mixed modes for $l=1$ and probably a
 few of the mixed modes for $l=2$ \citep{beck11}.


In order to clearly distinguish  the nature of the mixed modes, it is useful to study the behaviour of the displacement
eigenfunctions of the modes.
\begin{figure*}   
\centering  
\includegraphics[draft=false,scale=0.9]{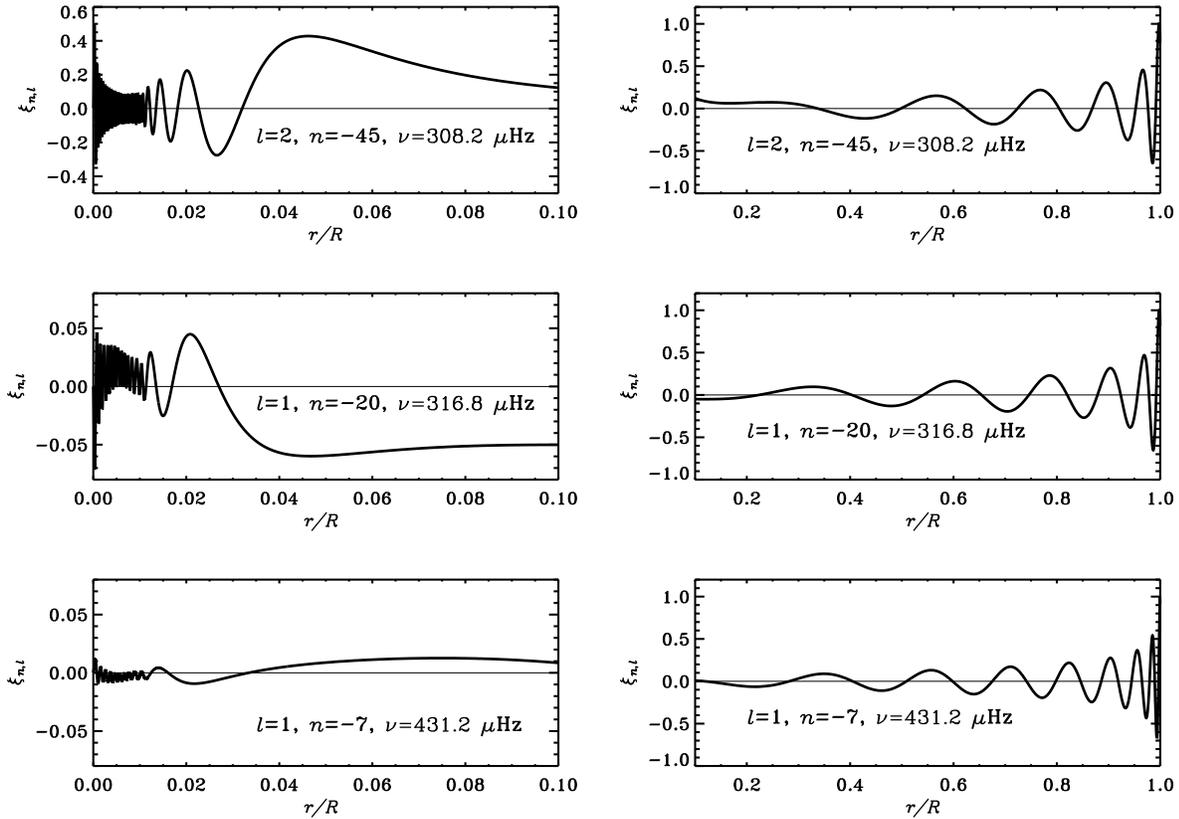}
\caption{Radial variations of eigenfunctions for three selected mixed modes calculated for Model~1 in Table~\ref{T.6}.
The left panels show the behaviour of the eigenfunction from the centre to $0.1\,R$. The right panels show the eigenfunctions from  $0.1\,R$ to the surface. The harmonic degree $l$, the radial order $n$ and 
the frequency  $\nu$ of the modes considered are indicated inside each panel.}
\protect\label{eigen}   
\end{figure*}   
Figure~\ref{eigen} shows the radial variation of the eigenfunctions of three mixed modes calculated for Model~1 of KIC~4351319 described in Table~\ref{T.6}.
The upper panels of Fig.~\ref{eigen} show the eigenfunction for a mixed mode with high inertia and hence with a predominant gravity nature, characterized by a quite large amplitude in the core.
The middle panels of Fig.~\ref{eigen} show the eigenfunction for a mixed mode with both gravity and pressure character. 
The lower panels of Fig.~\ref{eigen} show the eigenfunction for a nearly pure acoustic mode, with an inertia comparable to that of the radial modes: the amplitude in the core is negligible.

\subsection{Interpretation of the observed oscillation spectrum}

\begin{table*}
\caption{Relevant parameters,
the mass $ M$, the Age, the luminosity $L$, the effective temperature $T_{\rm eff}$, the initial hydrogen abundance $X_0$, the surface metallicity $Z_s$,
the surface radius $R$,
the location  
$r_{\rm cb}$ of the base of the convective envelope    
in units of $R$, the mixing-length parameter $\alpha$, the
diffusion coefficient $D_{\rm max}$ and the overshooting parameter $\alpha_{\rm ov}$  for a set of   
models of KIC~4351319.
}\label{T.6} 
\begin{center}
\begin{tabular}{ccccccccccccc
}\hline 
      &
$M/{\rm M_{\odot}}$ & 
 $  X_0$ &
$Z_s$ &
Age (Gyr) &
 $L/{\rm L_{\odot}}$ &    
 $T_{\rm eff} \; (\mathrm{K})$ &    
 $ R/{\rm R_{\odot}}$ &    
 $ \log g$ &
$r_{\rm cb}/R$&
$\alpha$&
$D_{\rm max}\, ({\rm cm^2\, s^{-1}})$ &
$\alpha_{\rm ov}$\\
 \hline   
$1$ &
$1.32$ &   
 $0.7$ &    
  $0.03$ &
 $ 5.29$ &   
  $5.25$ &    
 $4752$  &   
 $3.39$ &
$3.50$ &
$0.23$ &
$1.8$ &
No&
No\\
   
  $2$ &
$1.32$ &   
 $0.7$ &    
  $0.03$ &
 $ 5.28$ &   
  $5.23$ &    
 $4747$  &   
 $3.39$ &
$3.50$ &
$0.23$ &
$1.8$ &
$9000$& 
No\\

 $3$ &
$1.32$ &   
 $0.7$ &    
  $0.03$ &
 $ 5.28$ &   
  $5.22$ &    
 $4748$  &   
 $3.38$ &
$3.50$ &
$0.23$ &
$1.8$ &
40000& 
No\\
 $4$ &
$1.32$ &   
 $0.7$ &    
  $0.03$ &
 $ 5.27$ &   
  $5.28$ &    
 $4761$  &   
 $3.38$ &
$3.50$ &
$0.23$ &
$1.8$ &
No& 
$0.2$\\
 $5$ &
$1.28$ &   
 $0.7$ &    
  $0.03$ &
 $ 5.94$ &   
  $5.06$ &    
 $4734$  &   
 $3.35$ &
$3.49$ &
$0.22$ &
$1.8$ &
80000& 
No\\

\hline    
\end{tabular} 
\end{center}  
\end{table*}   
Among all the possible models, we selected a few models able to reproduce the set of all the observed frequencies of Table \ref{tab_final} and the observed values of the
 large and small separations.
The characteristics of some models that   
satisfy the observed constraints are given in   
Table~\ref{T.6}.   
According to the stellar evolution constraints, given
the match with the   
observed oscillation properties, and with the use of all the possible values   of mass and metallicity,   
our computations show that the age of KIC~4351319 is $(5.6\pm0.4)$\,Gyr, with a mass $M=1.30\pm0.03\, {\rm M}_\odot$, a radius
$R=3.37\pm0.03\, {\rm R}_{\odot}$ and a luminosity $L=5.1\pm 0.2\, {\rm L}_\odot$.\\ The values of mass and radius which have been obtained by direct modelling of the individual frequencies are better constrained and well compatible with those obtained by the scaling laws as described in Sect.~4.2.

\begin{figure*}
\centering  
\includegraphics[draft=false,scale=0.7]{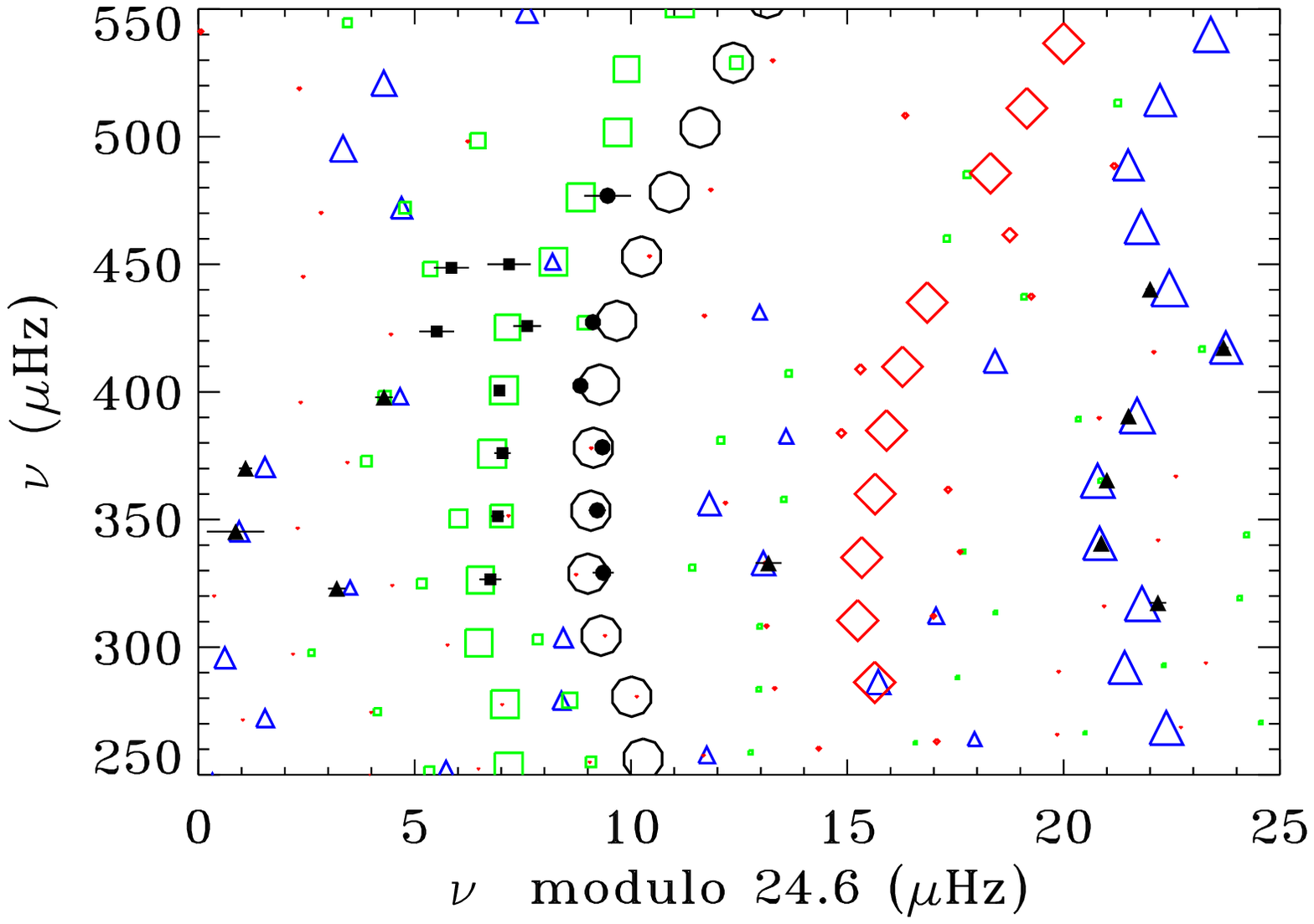}
\includegraphics[draft=false,scale=0.7]{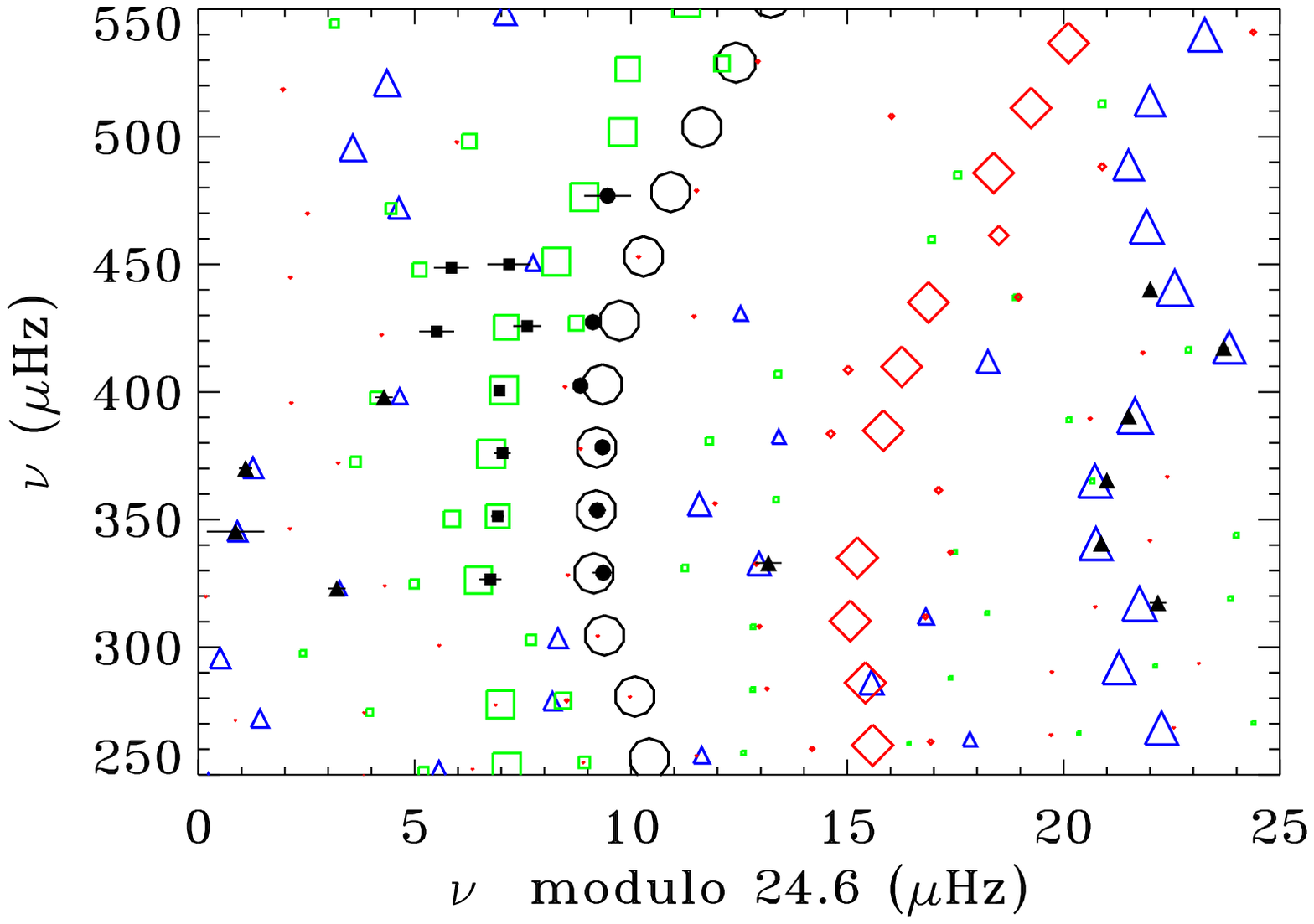}
\caption{Two \'echelle diagrams based on observed and computed frequencies,
plotted with $\Delta \nu=24.6 \, \mu\mathrm{Hz}$.
The upper panel shows results for Model~1 in Table~\ref{T.6}, while the lower panel shows results for Model~5 in
 Table~\ref{T.6}.
The open symbols represent computed frequencies, while the filled symbols
represent the observed frequencies of Table~\ref{tab_final}.
Circles are used for modes with $l=0$, triangles for $l=1$, squares for $l=2$, diamonds for $l=3$.   
The size of the open symbols indicates the relative surface amplitude of oscillation of the modes.}
\protect\label{echelle}   
\end{figure*}   

A detailed comparison between the theoretical oscillation spectra
for Model~1 and Model~5 of KIC~4351319 reported in Table \ref{T.6}
and the observed data is provided in the \'echelle diagrams
of Fig. \ref{echelle}.
 The size of the symbols is proportional to the theoretical oscillation    
amplitudes of p modes, relative to the amplitudes of radial   
modes with the same frequency (see Eq. \ref{ampl}).
 The results show,
as explained in previous sections, that the observed modes are
$l=0$ pure acoustic modes, and $l=1$ and $l=2$ nearly pure p modes and
some $l=1$ and $l=2$ g-p mixed modes.  Some non-radial modes with mixed
gravity-pressure character have an inertia so low as
 to propagate up to the surface and appear to behave
like solar-like oscillations.
Very few mixed modes, with a quite high inertia, keep their gravity character, although the combination with a p mode enhance their amplitude so that they can be observed at the surface and in the \'echelle diagram  appear to depart strongly from the regular solar-like pattern.\\
Pure g modes at lower frequencies, have not been detected with the present observations. Solar-like oscillations with $l=3$,
 although theoretically predicted,
 have not been
detected, probably due to geometrical reasons which produce the cancellation of the observed signal.

We found that there is a very good agreement between observed and theoretical frequencies. In particular we are able to reproduce all the $l=1$ mixed modes. 
The presence of mixed-modes with $l=2$ needs to be better investigated:
unfortunately none of
 our best models, shown in Fig. \ref{echelle}, is able to reproduce the $l=2$ mode with frequency $\nu=423.71\, \mu$Hz. In addition, the results
show that it is
 quite difficult
to determine which of Model~1 and Model~5 best reproduces the observations.
Perhaps the ambiguity will be solved by the identification
 of modes at lower
frequencies
and in particular if the reality of the
mode of $302.32\,\mu$Hz (see Sect.~4)
 will be confirmed or not.

Another important issue is related to the possibility of observing rotational splittings in this star. We estimate that the observed rotational velocity $v \sin i\simeq6\,{\rm km\,s^{-1}}$, quite high for this kind of objects, might produce a rotational splitting of $\sim 0.8\,\mu$Hz for the modes of $l=2$, $m=\pm2$ in case of rigid rotation. 
Such a rotational splitting is about twice the value of the frequency resolution of the $30\,d$ data set.
To check for the signature of rotation we have re-run the frequency determination now adding rotational split components for $l=1$ and $l=2$ modes. 
The rotation frequency and the inclination angle were allowed to vary in the range between $0-3\,\mu$Hz and $0^o-90^o$ respectively,
following the approach of \citet{gizon03}.
A comparison of the global likelihoods of the two models (with and without rotation) indicates that the rotation-free model is about 70 times more likely than the model including rotation.\\
Indeed, the presence of two modes closely spaced in frequency does not prove that we deal with rotational splitting although this possibility cannot be excluded.
Moreover, the two pairs of close frequencies of
$l=2$ suspected for evidence of rotational splittings
($448.65\, \mu$Hz with $449.98\,\mu$Hz and
$423.71\,\mu$Hz with $425.80\,\mu$Hz), appear to be not separated by a same amount in frequency, indicating that they cannot be both appearance of rotational splittings.

We finally notice that in order to reproduce the observed modes, we did not need to correct the frequencies for the
surface effects, as suggested by \citet{Kje08}, but the small departure of the $l=0$ modes at high frequency might be indicative of the need of such a correction.  In fact, the structure of the near-surface regions of the stars is quite uncertain: there are still substantial ambiguity in modelling the convective flux, defining an appropriate equation of state to describe the thermodynamic properties of the stellar structure, as well as in the treatment of non-adiabatic effects.
The conditions and limits  of the applicability of the surface effect are still unknown and will be the subject of future studies.\\
Since there is no evidence for fast rotation, we exclude any possible effect on the frequencies due to magnetic activity.

It was possible to determine the evolutionary state of this star through
  direct modelling, which places it in the hydrogen-shell-burning phase at
  the base of the ascending red-giant branch.  This star is substantially
  less evolved than the red giants, whose $l=1$ gravity-mode period spacings
  were measured by \citet{bedding11} and \citet{mosser11b}.  This
  is reflected in Fig. \ref{F.8} where we see that the spacing of the g-dominated
  modes (the upward sloping features) is comparable to the p-mode
  spacing. 
  This fact,
together with the very strong mode bumping, makes it is quite hard to see a
regular spacing in the l=1 modes of any single model. 
This makes it difficult to
  measure a clear g-mode period spacing.  However, we see from the list of
  observed modes in Table 3 that the consecutive $l=1$ modes have period
  spacings in the range 35--50\,s, which is consistent with a
  hydrogen-shell-burning star \citet{bedding11}.  Applying the method
  described by \citet{mosser11b} confirms a period spacing of about
  40$\pm$4\,s.

\subsection{Sharp features inside KIC~4351319}
\begin{figure}
\includegraphics[draft=false,scale=0.45]{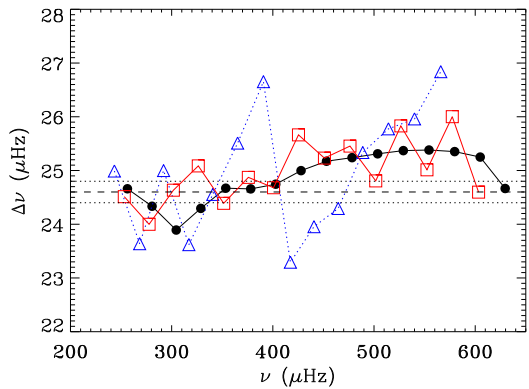}
\caption{Small oscillations in the
large frequency separation for acoustic modes of $l=0$ (black dots), $l=1$ (blue triangles) and $l=2$ (red squares)
 as calculated for Model~1 of Table \ref{T.6}. Observed large separation and its errors $\Delta\nu=24.6\pm0.2~\mu$Hz are indicated by dashed line and dotted lines respectively.} 
\label{oscsign}
\end{figure}

Figure \ref{oscsign} shows that a weak oscillatory motion is present in the calculated frequencies for the acoustic modes which follow closely the asymptotic law, namely the radial modes and the non-radial modes with very low inertia. This oscillatory signal, as explained in Sect. 3.3,  is produced only by the region of the second ionization of helium, which induces a local minimum in the
 first adiabatic coefficient $\Gamma_1$ and hence a sharp variation in the sound speed.

\begin{figure}
\centering
\includegraphics[draft=false,scale=0.45]{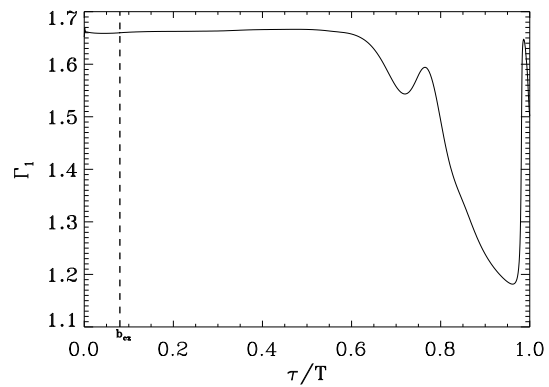}
\caption{First adiabatic exponent as function of the acoustic radius for Model~1 of Table \ref{T.6}. The region where helium undergoes its second ionization corresponds to the second minimum in $\Gamma_1$. The location of the base of the convective envelope is also indicated.}
\protect\label{Gamma1}
\end{figure}

Figure \ref{Gamma1}, which shows the behaviour of the first adiabatic coefficient as function of the acoustic radius 
$t=\int_0^{r_1}{\rm d}
r/c(r)$, indicates that in the models of KIC~4351319
  the second helium ionization zone occurs at about $t_{\rm HeII}/T\simeq 0.7$, while
the base of the convective envelope is located at $t_{\rm b_{cz}}/T\simeq 0.08$,
 too deeply to produce an effective signal into the solar-like frequencies \citep{miglio10}. 
 The properties of the helium ionization zone, once determined from the oscillation frequencies, may be used to constrain the structure of the star, in particular the envelope helium abundance.
Further studies and developments are required in order to determine
how a different content of He and a different equation of state for the computation of the models will result in a different oscillatory behaviour in the theoretical frequencies.

\begin{figure*}
\centering  
\includegraphics[draft=false,scale=0.7]{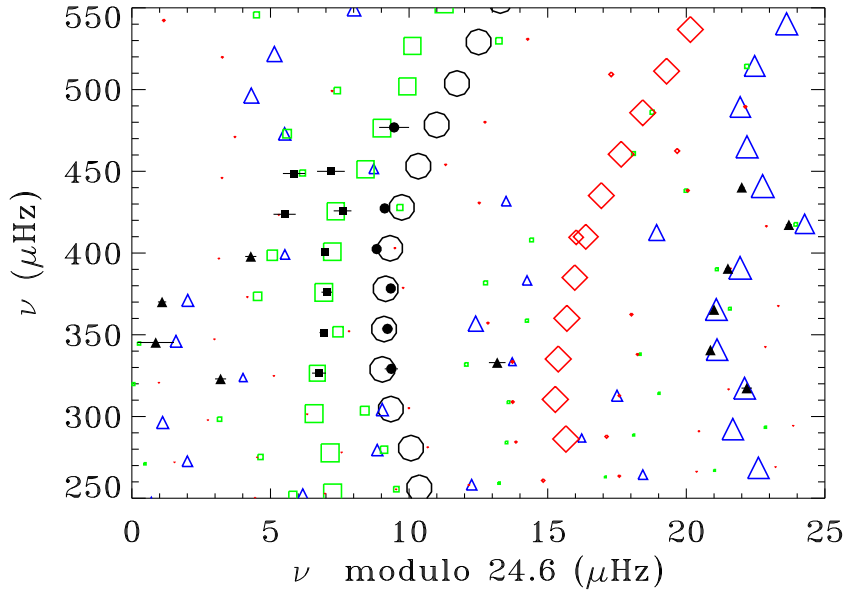}
\caption{\'Echelle diagram based on observed and computed frequencies. 
plotted with $\Delta \nu=24.6 \, \mu\mathrm{Hz}$. 
The open symbols show computed frequencies for Model~2 calculated with turbulent diffusion and described in Table    
\ref{T.6}.
Circles are used for modes with $l=0$, triangles for $l=1$, squares for $l=2$, diamonds for $l=3$.   
The size of the open symbols indicates the relative surface amplitude of oscillation of the modes.}
\protect\label{echelle2}   
\end{figure*}

Secondly, we have also considered the possibility to isolate the signal coming from the base of the convective envelope, by looking at the observed modes which
penetrate deeply inside the star, namely the mixed modes with a high inertia.
This can be studied following \citet{miglio08}, by comparing models  modified by adding turbulent diffusion or overshooting at the base of the convective envelope.
The results have shown that the inclusion of turbulent diffusion, or overshooting, below the convective envelope produces a displacements only in the mixed modes for $l=1$ and $l=2$ modes, while all the other frequencies are not modified as shown in Fig. \ref{echelle2}. Thus, although we have not observed high-order g modes, we can still use the mixed modes to probe qualitatively the interior of this star. However, it is impossible to distinguish the effect on the frequencies due to the inclusion of diffusion from that of overshooting. 

\section{Conclusion}

KIC~4351319 has been observed for 30 days by the {\it Kepler} satellite.
These observations have yielded  a clear detection of
25 modes identified with harmonic degrees $l=0,1,2$ between $300$ and $500\,\mu$Hz with a large separation $\Delta \nu_{0} = 24.6 \pm 0.2\, \mu$Hz and 
 a small separation
$\delta \nu_{02} =
2.2 \pm 0.3 \,\mu$Hz respectively.

The oscillation spectrum of this star is characterized by the presence of a well defined solar-like oscillations pattern due to
 radial acoustic modes and non-radial nearly pure p modes equally spaced in frequencies. 
The nearly pure p modes have mixed
gravity-pressure character with an inertia so low as
 to propagate up to the surface and appear to behave
like solar-like oscillations, although they penetrate deeply to the core.

The oscillation spectrum also showed evidence for the presence of mixed modes
with strong gravity-mode character. Those modes sound the internal region of this star and, besides constraining the stellar age, carry tight information about the location of the convective envelope
and the condition in the core. This enabled to define characteristics of this star with an accuracy which cannot be reached by the use of only the global asteroseismic parameters, $\Delta\nu$ and $\nu_{\rm \max}$. 

Thus, in this paper we have addressed the problem of identifying the structural properties and the  
evolutionary state of KIC~4351319 by using the results of the analysis of the oscillation spectrum
 and
the atmospheric parameters provided by supplementary ground-based spectroscopic observations.

Detailed modelling of this star, in the attempt to match both the asteroseismic and spectroscopic constraints, allowed us to determine the main parameters with an unprecedent level of precision for a red-giant star, with uncertainties of $2\,\%$ for mass, $7\,\%$ for age, $1\,\%$ for radius, and  $4\,\%$ for luminosity.
In the end, we are able to conclude that this star is in the red-giant phase of the evolution, with a mass $M=(1.30\pm0.03)\, {\rm M}_\odot$, an age of      
 $5.6\pm0.4$~Gyr, a radius
$R=3.37\pm0.03\, {\rm R}_{\odot}$ and a luminosity $L=5.1\pm 0.2\, {\rm L}_\odot$.
The uncertainties obtained for the stellar parameters come from the degeneracy of the models describing the 
stars in the red-giant phase of the evolution.

 Only the detection of g modes, which better sound the condition of the core,
could improve the results obtained by the present analyses, while strong theoretical efforts
need to be done in order to achieve a better description of 
the stellar structure. 

It is clear that KIC~4351319 represents an excellent candidate for
long-term observations, which have been already scheduled for {\it Kepler} future runs.
This will allow us to open a window on the understanding of this phase of the evolution and to
 unveil all the raised questions including those relative to the identification of the mixed modes, to the signature of the ionization of the He and to the presence of diffusion or overshooting
below the convective envelope.
In particular, the possibility to detect rotational splittings in KIC~4351319 appears very fascinating and represents an unique opportunity for sounding the internal dynamics of a red-giant star.

\section*{Acknowledgments}
Funding for this mission is provided by NASA's Science Mission Directorate. We thank the entire Kepler team for the
development and operations of this outstanding mission.\\
The work presented here is also based on observations obtained with the Harlan J. Smith Telescope at McDonald Observatory, Texas.\\
SH acknowledges financial support from the UK Science and Technology Facilities Council (STFC).
SH acknowledges financial support from the Netherlands Organization for Scientific Research (NWO).\\
KU acknowledges financial support by the Deutsche Forschungsgemeinschaft (DFG) in the
framework of project UY 52/1-1.\\
TK is supported by the Canadian Space Agency and the Austrian Science Fund (FWF P22691-N16)

\appendix
\section{Maximum Likelihood Estimators (MLE)}
According to this method, the p-mode parameters are estimated by finding the best fit between a modelled and
the observational power density spectrum by using a maximum-likelihood technique \citep{anderson90}.
Two different models have been assumed:
\subsection{Lorentzian with convolution of the window function}
The fitting of the non-oversampled power spectrum has been done in two steps. First the background, for which we use a power law fit to account for long term effects, such as granulation, rotation or activity related to spots, is fitted.
Then, we computed a global maximum likelihood fit \citep{anderson90} to all oscillation modes, where the final model $M$ is a convolution of the model with the power spectrum of the observed window function of the data set, normalised to unit total area (see Eq.~\ref{model}). This takes the redistribution of power caused by gaps in the data into account.
\begin{equation}
M=\left(\sum_j \frac{H_j}{1+((\nu_{\rm cen,\it j} -\nu)/B_j)^2}\right) \ast \rm window
\label{model}
\end{equation}
In this model the previously determined background is kept fixed, while each of the $j$th oscillation peak is fitted with the height ($H$), the central frequency ($\nu_{\rm cen}$) and HWHM ($B$) and the noise level as free parameters.

The selection of the oscillation modes is based on a statistical test of the binned power spectrum. For this test we bin the power spectrum over intervals of three frequency bins. Then we compute the probability of the power in the binned power spectrum to be due to noise. This probability is computed using a $\chi^2$ distribution in which we take the width and number of the bins into account in the degrees of freedom. Frequencies at which the probability of the power not being due to white noise is larger than 95\% are selected as candidate oscillation frequencies. After performing the fit, we also verify that in the ratio of the observed to the fitted spectra no prominent peaks are left. Therefore we compute the relative height for the investigated frequency range for which the probability of observing at least one spike with this height due to noise is 10\%, following the formulation by \citet{chaplin02} and references therein.\\
This is also described in \citet{hekker10a}.

\subsection{Sum of Lorentzians or sinc}
 If the modes are resolved (i.e. the mode width is larger than the frequency resolution)
we chose to model the oscillation spectrum by a sum of Lorentzians plus a background model
that counts for the signal which is not due to the p modes such as the instrumental noise or stellar background. In the frequency range considered here, the background has been
modelled by a straight line (i.e. $a + b \nu$). When the mode width is of the order of
the frequency resolution we change the sum of Lorentzians by a sum of sinc.

The modelled power spectrum used to match the data is:

\begin{equation}
P(\nu_k) = \sum_{n=1}^{Q} \cal M\rm(n,\nu_k)  +B(\nu_k),
\end{equation}
where $Q$ is the number of oscillation modes, $n$ the radial order, $\nu_k$ the Fourier
frequencies, and   $B(\nu_k)$ the background
noise in the power spectrum, modelled as a straight line, i.e. $a + b\nu_k$ with $a$ and
$b$, two constants and free parameters of the fit. $\cal M\rm(n,\nu_k)$ is a Lorentzian
when the mode width is larger than the frequency resolution:
\begin{equation}
\cal M\rm(n,\nu_k) = H_n \frac{1}{1 + \left(\frac{2(\nu_k-\nu_n)}{\Gamma_n}\right)^2}
\end{equation}
where $H_n$ is the height of the
Lorentzian profile, $\nu_n$
is the oscillation mode frequency and $\Gamma$ is the mode line width
(FWHM) with $\Gamma =1/(\pi \tau)$, $\tau$ being the mode lifetime.

When the modes are not resolved, $\cal M\rm(n,\nu_k)$ is a sinc function:
\begin{equation}
\cal M\rm(n,\nu_k) = H_n sinc^2(\Pi T (\nu_k-\nu_n))
\end{equation}
where $T$ is the total length of the observing run.

The power spectrum is fitted ``globally'' over a frequency range corresponding to the
detected excess power.
The free parameters of the fitting process are:

\begin{itemize}

\item  the height of the
Lorentzian profile, $H_n$, $n$ being the radial order. A single height parameter is
fitted per mode.

\item  the frequency for each mode, $\nu_n$.

\item the line width for each mode
(FWHM) with $\Gamma_n = 1/(\pi \tau_n)$, $\tau_n$ being the mode lifetime. A single width has been considered.

\item the parameters $a,b $ describing the background model as mentioned above.

\end{itemize}

The mode-parameter 1$\sigma$ error bars are derived from the Hessian
matrix.
No oversampling has been used in the computation of the power spectrum for
the fitting procedure in
order to minimize the correlation of the points.
The degree of each mode is identified using \'echelle diagrams.

\section{Bayesian MCMC method}
The p-mode parameters are estimated by fitting a sequence of Lorentzian profiles to the power density spectrum, where we use the previously determined stellar activity and granulation signal (Eq. 4 without the Gaussian) as a fixed background. The visually identified mode profiles are parameterized by their central frequencies, mode heights, and lifetimes. As it can be seen in Fig. 2, the mode profiles are quite narrow. 
To prevent the algorithm from over-fitting the data and to keep the number of free parameters to a reasonable amount we use two lifetime parameters, one for l=0 and 2 modes and one for l=1 modes. Furthermore, we assume the individual mode heights to follow a Gaussian envelope parameterized by a single central frequency and width but with individual heights for each mode degree. For the 25 modes in Tab. 3, this results in a total of 32 free parameters. For the fit we use a Bayesian MCMC algorithm 
\citep{Gruberbauer09} that delivers probability density functions for all fitted parameters and their marginal distributions, from which we compute the most probable values and their 1$\sigma$ uncertainties. During the fit the mode frequency parameters are allowed to vary independently within $\pm$2$\mu$Hz around the value inferred from a visual inspection of the spectrum. The lifetimes are kept between 1 and 50 days. Whereas the centre and width of the Gaussian envelope are kept within 0.9 and 1.1 times $\nu_{\mathrm{max}}$ and $\sigma$ (Eq.4), respectively, the height for each mode degree is allowed to vary independently between 0 and 2 times the highest peak in power density spectrum. 
\end{document}